\documentclass[english,usenatbib]{mn2e}
\pdfoutput=1
\usepackage[]{fontenc}
\usepackage[latin1]{inputenc}
\usepackage{amsmath}
\usepackage{graphicx}
\usepackage{amssymb}
\usepackage[authoryear]{natbib}
\usepackage{longtable}

\onecolumn

\newcommand{\bfx}{{\bf x}}
\newcommand{\bfy}{{\bf y}}

\newcommand{\eq}[1]{\mbox{Eq. #1}}
\newcommand{\fig}[1]{\mbox{Fig. #1}}

\title[Astrophysical WPMHD]{Astrophysical Weighted Particle Magnetohydrodynamics}

\author[E. Gaburov, K. Nitadori]{Evghenii Gaburov$^{1}$,  Keigo Nitadori$^{2}$\\
$^1$ Leiden Observatory, Leiden University, the Netherlands\\
$^2$ RIKEN, Tokyo, Japan
}

\begin{document}

\maketitle

\begin{abstract}
This paper presents applications of weighted meshless scheme for conservation laws to the
Euler equations and the equations of ideal magnetohydrodynamics. The divergence constraint
of the latter is maintained to the truncation error by a new meshless divergence cleaning
procedure. The physics of the interaction between the particles is described by an
one-dimensional Riemann problem in a moving frame. As a result, necessary diffusion which
is required to treat dissipative processes is added automatically. As a result, our scheme
has no free parameters that controls the physics of inter-particle interaction, with the
exception of the number of the interacting neighbours which control the resolution and
accuracy. The resulting equations have the form similar to SPH equations, and therefore
existing SPH codes can be used to implement the weighed particle scheme. The scheme is
validated in several hydrodynamic and MHD test cases. In particular, we demonstrate for
the first time the ability of a meshless MHD scheme to model magneto-rotational
instability in accretion disks.
\end{abstract}

\section{Introduction}\label{sect:introduction}
Computational magnetohydrodynamics (MHD) remains an important tool to understand complex
behaviour of astrophysical plasmas. While many methods have been developed to solve
equations of ideal MHD of Eulerian (cartesian) meshes, few successful Lagrangian meshless
formulations exists. The latter, however, are desired for problems which lack particular
symmetries, cover many length-scales or require adaptivity, for example stellar collisions
star or star cluster formations.

Smoothed particle hydrodynamics (SPH) proved to be a successful Lagrangian meshless scheme
to solve equations of fluid dynamics in variety research fields
\citep{0034-4885-68-8-R01}. Despite its limitations, it has been also successfully used in
wide range of astrophysical problems, including evolution of gaseous disks around black
holes or stars, star formation, stellar collisions, and cosmology. Because of its
simplicity and versatility, several attempts, albeit with limited success, have been made
to include magnetic fields into SPH, thereby formulating smoothed particle MHD, or SPMHD
for short \citep{2005MNRAS.364..384P, 2006ApJ...652.1306B, 2007MNRAS.379..915R}. It soon
became clear that SPMHD equations where plagued with two main problems: tensile
instability and maintenance of the divergence constraint, $\nabla\cdot{\bf B} = 0$. The
former is a general problem of SPH, namely the equations are unstable to tensile stresses
\citep{Swegle1995123, 2000JCoPh.159..290M}. In case of SPMHD, this instability manifest
itself as particle clumping in the regions where magnetic pressure dominates gas pressure,
and therefore rendering the simulation of strongly magnetised plasma unfeasible. It has
been shown that the equations can be stabilised by either adding short-range repulsive
forces between particles (e.g. \citealp{2004MNRAS.348..123P}), or sacrificing momentum
conservation (e.g. \citealp{2005MNRAS.364..384P}). Alternatively,
\cite{2001ApJ...561...82B} showed that the stability of SPMHD equations can also be
achieved by adding a source term proportional to the divergence of magnetic field to the
momentum equation. While these approaches appear to remove the tensile instability, the
divergence constraint still remains an issue in these SPMHD
formulations. \cite{2007MNRAS.379..915R} were able to formulate manifestly divergence-free
SPMHD equations which appears to work for a wide range of problems
(e.g. \citealp{2006Sci...312..719P}, \citealp{2008MNRAS.385.1820P}). However, they solve a
limited form of the induction equation which only permits topologically trivial field
configurations and is unable to model more complicated MHD phenomena, such as
magneto-rotational instability. \cite{2009MNRAS.398.1678D} were also able to formulate
stable SPMHD equations by using a combination of several techniques, such as the addition
of the source term proportional to the magnetic divergence to the momentum equation,
artificial dissipation and smoothing of the magnetic field, and modification of the
induction equation. This approach introduces several free parameters, which the authors
were able to constrain by fitting their results to the solutions of several shock tube
problems computed with conservative Eulerian MHD schemes. Despite the progress in this
field, the difficulties associated with formulating consistent SPMHD equations advocate
the need of the alternative approaches to formulate meshless Lagrangian MHD schemes. In a
gradient particle magnetohydrodynamics (GPMHD, \citealp{2003ApJ...595..564M,
  2005prpl.conf.8461M}), the equations of ideal MHD were discretised on a set of particles
by fitting a second or forth order polynomial into the data in order to obtain first and
second order derivatives of the desired quantities. However, such approach still requires
addition of artificial diffusion to the resulting equations to model dissipative and
resistive processes across discontinuities, and this introduces additional free parameters
which control dissipative processes. In addition, the local conservation, for example
third Newton law, is satisfied to the truncation error only. This is a potential source of
error for the problems with interacting strong shock waves, since the truncation error
across a shock wave is always of order of unity. Nevertheless, GPMHD appears to be a
viable, albeit noticeably more complex, alternative for SPMHD, at least for subsonic and
weakly supersonic flows.

Alternatively, one may attempt to utilise a Godunov approach. It has been shown that most
of, if not all, SPH limitation can be removed by solving a hydrodynamic Riemann problem
between each pair of interacting particles to obtain pressure forces, instead of computing
separately pressure forces and artificial viscosity terms \citep{2002JCoPh.179..238I,
  2003MNRAS.340...73C, 2010MNRAS.403.1165C}. In such Godunov SPH (GSPH) formulations, the
necessary mixing and dissipation is included into underlying Riemann problem which
described the physics of the interaction. Borrowing these ideas, it is therefore
conceivable that Godunov-like meshless MHD formulation will eliminate the problems that
plague SPMHD formulation, thereby permitting formulation of consistent Lagrangian meshless
MHD scheme.

In this paper we formulate a weighted particle MHD scheme. Our scheme is based on a meshless
discretisation of the conservation law equations, which was pioneered by \cite{VILA:1999},
and in Section \ref{sect:methods} we present a heuristic derivation of the meshless
conservative equations. We give the implementation details of our scheme in Section
\ref{sect:Implementation}. In Section \ref{sect:ideal_hd} we present applications to the
equation of ideal hydrodynamics, and in Section \ref{sect:ideal_mhd} we show how our
scheme can be applied to the equations of ideal MHD. In Section \ref{sect:results} we
validate our meshless MHD scheme on several test problems, and finally, we present our
conclusions in Section \ref{sect:discussion}.

\section{Methods}\label{sect:methods}

\subsection{Meshless equations for conservation laws}\label{sect:meshless_equations}

In what follows, we present a heuristic derivation of meshless discretisation of a scalar
conservation law. The readers interested in a rigorous mathematical formulation
supplemented with convergence theorems are referred to the original papers by
\cite{VILA:1999} and \cite{1404789, 1404790}. Following these works, a weak solution to a
scalar conservation law
\begin{equation}
  \frac{\partial u}{\partial t} + \nabla\cdot\left({\bf F} + {\bf a}u \right) = S,
\end{equation}
is defined by
\begin{equation}
  \int \left( u(\bfx,t)\dot\varphi + {\bf F}(u,\bfx,t)\cdot\nabla\varphi +
  S(\bfx,t)\varphi \right)\,d\bfx\,dt = 0.
  \label{eq:weak_formulation}
\end{equation}
Here, $u(\bfx,t)$ is a scalar field, $S(\bfx, t)$ is its source, ${\bf F}(u,\bfx,t)$ is
its flux in a frame moving with velocity ${\bf a}(\bfx, t)$, and the integral is carried
out over all space-time domain of a problem at hand. The function
$\varphi\equiv\varphi(\bfx,t)$ is an arbitrary differentiable function in space and time,
$\dot\varphi = \partial\varphi/\partial t + {\bf a}(x,t)\cdot\nabla\varphi$ is an
advective derivative, and ${\bf a}(\bfx, t)$ is an arbitrary smooth velocity field which
will describe motion of particles. This integral is discretised on a set of particles with
coordinates $\bfx_i$ with the help of a partition of unity
\begin{equation}
  \psi_i(\bfx) = w(\bfx)W(\bfx-\bfx_i,h(\bfx)),
  \label{eq:partiaion_of_unity}
\end{equation}
where, $w(\bfx)^{-1} = \sum_j W(\bfx-\bfx_j,h(\bfx))$ is an estimate of the particle
number density and the sum is carried out over all particles, $h(\bfx)$ is a smoothing
length, and $W(\bfx, h)$ is a smoothing kernel\footnote{In this paper we use cubic-spline
  smoothing kernel, which is commonly used in SPH.} with a compact support of size $h$; in
what follows we assume that the kernel is normalised to unity. Inserting $1 =
\sum_i\psi_i(\bfx)$ into an integral of an arbitrary function, we obtain
\begin{equation}
  \int f(\bfx)\,d\bfx = \sum_i \int f(\bfx)\psi_i(\bfx)\,d\bfx \approx \sum_i f_i \int \psi_i(\bfx)\,d\bfx
  \equiv \sum_i f_i V_i,
  \label{eq:discretisation}
\end{equation}
where $V_i = \int \psi_i(\bfx)\,d\bfx$ is the effective volume of a particle $i$, and in the
third term we use first-order Taylor expansion of $f(\bfx)$. In principle, a higher order
discretisation is also possible, but for the purpose of this work such an one-point
quadrature is sufficient; in fact, on a regular distribution of particles this
discretisation is second order accurate (c.f. \S\ref{sect:regularity}). Application of
this discretisation to \eq{\ref{eq:weak_formulation}} gives
\begin{equation}
  \sum_i \int \left(
  V_i u_i \dot\varphi_i + 
  V_i F^\alpha_i (D^\alpha\varphi)_i +
  V_i S_i \varphi_i
  \right) = 0.
  \label{eq:weak_discrete}
\end{equation}
Here, the Einstein summation is assumed over Greek indexes, which refer to the components
of a vector. Also, the gradient of a function $\varphi$ at $i$-particle location,
$(\nabla\varphi)^\alpha_i$ is replaced by its discrete version $(D^\alpha\varphi)_i$, and
this, for example, can be computed with SPH estimates of a gradient. However, a much
better approach is to employ a more accurate meshless gradient estimate suggested by
\cite{1404789}. They showed that a second order accurate meshless partial derivative is
given by the following expression
\begin{equation}
  (D^\alpha f)_i = 
  \sum_j(f_j - f_i)B^{\alpha\beta}_i \Delta x_{ij}^\beta\psi_j(\bfx_i) \equiv
  \sum_j(f_j - f_i)\psi^\alpha_j(\bfx_i),
  \label{eq:strong_derivative}
\end{equation}
where $\psi^\alpha_i(\bfx_i) = B_i^{\alpha\beta}\Delta x_{ij}^\beta\psi_j(\bfx_i)$, $\Delta
x_{ij}^\alpha = (\bfx_j - \bfx_i)^\alpha$ and $B_i^{\alpha\beta} = (E^{\alpha\beta}_i)^{-1}$ is
a renormalisation matrix defined by its inverse
\begin{equation}
  E_i^{\alpha\beta} = \sum_j \Delta x_{ij}^\alpha \Delta x_{ij}^\beta \psi_j(\bfx_i).
  \label{eq:renorm_matrix}
\end{equation}
Finally, integrating the first term by parts with assumption that $\phi$ vanishes at
boundaries, and applying the following rearrangement to the second term
\begin{equation}
  \sum_i V_i F_i^\alpha(D^\alpha\varphi)_i = +\sum_{i,j} V_i
  F_i^\alpha\varphi_j\psi^\alpha_j(\bfx_i) - \sum_{i,j}V_i
  F_i^\alpha\varphi_i\psi^\alpha_j(\bfx_i)  = -\sum_i\varphi_i\sum_j\left(V_i
  F_i^\alpha\psi^\alpha_j(\bfx_i) - V_j F_j^\alpha \psi^\alpha_i(\bfx_j)\right) 
\end{equation}
permits separation of $\varphi$ from the rest
\begin{equation}
  \int\,dt\,\sum_i\varphi_i\left(
  - \frac{d}{dt}(V_i u_i) 
  - \sum_j\left[V_i F_i^\alpha\psi_j^\alpha(\bfx_i) - V_j F_j^\alpha\psi_i^\alpha(\bfx_j)\right] 
  + V_i S_i
  \right) = 0.
\end{equation}
The above is true for an arbitrary function $\varphi$ if the expression in brackets
vanishes, namely
\begin{equation}
  \frac{d}{dt}(V_i u_i) + 
  \sum_j\left[V_i F_i^\alpha\psi_j^\alpha(\bfx_i) - V_j F_j^\alpha\psi_i^\alpha(\bfx_j)\right] 
  = V_i S_i.
  \label{eq:meshless_conservation_law}
\end{equation}
The extension of this equation to a general vector field ${\bf u}$ is straightforward:
this equation is applied to each component of the field. These equations are similar to
SPH equations, with the difference that the physics of the particle interaction is hidden
in the fluxes and source terms. Indeed, if one uses the fluxes of ideal Lagrangian
hydrodynamics, the equation of motions similar to SPH can be derived. As with SPH, such
equations do not include dissipative processes, and therefore must be augmented with
explicit diffusive terms, namely in the form of artificial viscosity, conductivity and
resistivity.

However, the power of the new scheme becomes apparent with the realisation that one can
utilise the fluxes produced by the solution of an appropriate Riemann problem between
particles $i$ and $j$, which automatically include necessary dissipation. Defining such a
flux as $\bar F_{ij}^\alpha$, and setting $F_i^\alpha = F_j^\alpha = \bar F_{ij}^\alpha$
gives
\begin{equation}
  \frac{d}{dt}(V_i u_i) + \sum_j \bar F_{ij}^\alpha 
  \left[V_i\psi_j^\alpha(\bfx_i) - V_j\psi_i^\alpha(\bfx_j)\right] = S_i V_i.
  \label{eq:meshless_flux}
\end{equation}
Finally, defining a vector $n^\alpha_{ij} = V_i\psi_j^\alpha(\bfx_i) - V_j\psi_i^\alpha(\bfx_j)$
and $\hat{\bf n} = {\bf n}/|{\bf n}|$, the equations take the following form
\begin{equation}
  \frac{d}{dt}(V_i u_i) + \sum_j (\bar{\bf F}_{ij}\cdot \hat{\bf n}_{ij})|{\bf n}_{ij}| =
  S_i V_i.
  \label{eq:meshless_1D}
\end{equation}
It becomes clear, only the projection of the flux on the direction of the vector ${\bf
  n}_{ij}$ is required, and therefore for a wide range of problems the flux can be
obtained by solving an appropriate 1D Riemann problem in a frame moving with mean velocity
of the two particles, ${\bf a}_{ij}$.

\subsection{Linear monotonic reconstruction}\label{sect:linear_reconstruction}
The one dimensional flux in \eq{\ref{eq:meshless_1D}} is naturally computed at the
midpoint between particles $i$ and $j$, i.e. at $\bfx_{ij} = (\bfx_i + \bfx_j)/2$. To achieve
higher than the first order accuracy, it is necessary to linearly reconstruct left and
right states of the Riemann problem to this location. The reconstruction step should in
principle be done in characteristic variables, however for the second and third order
schemes this can be done in primitive variables, ${\bf w}$, as well. In the case of MHD,
the latter are density $\rho$, pressure $p$, velocity ${\bf v}$ and magnetic field ${\bf
  B}$. An approximation of ${\bf w}_{ij;i}$ of an $i$-paritcle state at $\bfx_{ij}$ is given
by a first-order Taylor expansion from the point $\bfx_i$:
\begin{equation}
  {\bf w}_{ij;i} = {\bf w}_i + \tau_i (\bfx_{ij} - \bfx_i)^\alpha (D^\alpha{\bf w})_i,
  \label{eq:linear_reconstruction}
\end{equation}
where $(D^\alpha{\bf w})_i$ is a gradient estimate of the primitive variables computed
with \eq{\ref{eq:strong_derivative}}, and $\tau_i$ is a vector of limiting functions which
is required in order to assure non-oscillatory reconstruction \citep{2004ApJS..151..149B}
\begin{equation}
  \tau_i = \min\left[1, \kappa \min\left( \frac{{\bf w}_{i, {\rm ngb}}^{\rm max} - {\bf
        w}_i}{{\bf w}_{i,{\rm mid}}^{\rm max} - {\bf w}_i}, \frac{{\bf w}_i - {\bf w}_{i,
        {\rm ngb}}^{\rm min}}{{\bf w}_i - {\bf w}_{i,{\rm mid}}^{\rm min}} \right)
    \right].
  \label{eq:limiter}
\end{equation}
Here, ${\bf w}_{i, {\rm ngb}}^{\rm min}$ and ${\bf w}_{i, {\rm ngb}}^{\rm max}$ are the
minimal and maximal ${\bf w}$ respectively over all neighbours that particles $i$
interacts with, and ${\bf w}_{i, {\rm mid}}^{\rm min}$ and ${\bf w}_{i, {\rm mid}}^{\rm
  max}$ are the minimal and maximal ${\bf w}$ resulted from the reconstruction in
\eq{\ref{eq:linear_reconstruction}} for each of these neighbours.  The scalar constant
vector $\kappa$ should have values between $0.5$ and $1.0$ in order to achieve second
order of accuracy. Following suggestions of \cite{2004ApJS..151..149B}, value $0.5$ should
be used for both pressure and velocity, and $1.0$ for both density and magnetic field;
however, we find no problems while using $1.0$ for all fluid quantities.
The frame velocity at $x_{ij}$ is approximated as ${\bf a}_{ij} = ({\bf a}_i + {\bf
  a}_j)/2$. Finally, the reconstructed left and right states, the frame velocity ${\bf
  a}_{ij}$, and the unit vector ${\bf n}_{ij}$ are used to obtain the flux from the 1D
Riemann problem.

Instead of a linear, a piecewise parabolic reconstruction can also be used to achieve
third-order spatial accuracy. In Appendix \ref{sect:ppm_reconstruction}, we describe
parabolic reconstruction of a scalar field $q(\bfx,t)$. Due to large operation count, this
reconstruction is presented only for purpose of completeness, and in the test that will
follow later, only linear reconstruction is used.

\section{Implementation}\label{sect:Implementation}
\subsection{Smoothing length}\label{sect:Smoothing_length}
The smoothing length in our scheme is a property of the particle distribution and, in
contrast to SPH, does not depend on the fluid state. In principle, a constant smoothing
length can be used throughout the whole space and time domain. In practice however this
cause difficulties due to the possible development of wide range in particle number
densities as the simulation progresses. Similarly to SPH, this can lead to under- or
oversampling in low and high particle density regions respectively. The approach used here
is inspired by conservative SPH formulations \citep{2002MNRAS.335..843M,
  2002MNRAS.333..649S}. The idea is to constrain the smoothing length of a particle $i$,
$h_i = h(\bfx_i)$, to the particle number density at this location, $n_i = n(\bfx_i)$,
i.e.
\begin{equation}
  C n_i h_i^D = N_{\rm ngb}.
  \label{eq:constraint_h}
\end{equation}
This tends to maintain approximately $N_{\rm ngb}$ number of neighbours for each particle;
here $C = 1$, $\pi$ and $4\pi/3$ for $D=1$, $2$ and $3$ dimensions respectively, and
$n(\bfx_i) = 1/w(\bfx_i)$, where $w(\bfx_i)$ is defined in
\eq{\ref{eq:partiaion_of_unity}}. As in conservative SPH equations, $h_i$ is obtained by
iteratively solving \eq{\ref{eq:constraint_h}}, for example via Newton-Raphson method
(e.g.  \citealp{1992nrca.book.....P}). One might be also tempted to use the continuity
equation
\begin{equation}
  \frac{dh(\bfx)}{dt} = \frac{h(\bfx)\nabla\cdot{\bf a}}{D},
\end{equation}
to compute time evolution of the smoothing length from its initial value. This, however,
is undesirable for two main reasons: a) the result depends on the functional form of the
divergence operator, and b) in discontinuous flows the $\nabla\cdot{\bf a}$ may be
undefined at some points, which can result in unexpected behaviour. As a result, in our
tests we chose to iteratively solve \eq{\ref{eq:constraint_h}}, but we use the
differential form to predict $h(\bfx)$ as a first guess to an iterative solver.

Finally, knowledge of smoothing length permits calculation of the rest of geometric
quantities, such as effective volume of a particle, $V_i$. It is possible to use numerical
quadrature to evaluate $\int \psi_i(\bfx) \,d\bfx$ with a desired accuracy, however we find that
defining $V_i = w(\bfx_i)$ works fine for our purpose, and therefore we decided not to
perform more accurate volume estimates.

\subsection{Particle regularity}\label{sect:regularity}
Particle regularity is an important aspect of the scheme. If particles are randomly
sampled within a domain, there is non-zero probability that particle's smoothing length,
$h$, will differ significantly from the average $h$ in its neighbourhood.  Furthermore,
the resulting $h$-distribution will not be a smooth function of position, and therefore
will not be differentiable. This will break the approximation which lead to
\eq{\ref{eq:meshless_conservation_law}}. Namely, the variation of $h$ within the neighbour
sphere will be large enough that the estimate in \eq{\ref{eq:discretisation}} will result
in intolerable errors which produces unexpected behaviour, such as negative values of
density or pressure. To avoid these situations, the particle distribution must be first
regularised. If the initial particle distribution is regular, it will maintain its
regularity during the simulation except in the regions where particle velocity field,
${\bf a}$, is discontinuous, e.g. across shock waves \citep{VILA:1999, 1404789}. The
criteria which determines regularity of the particle distribution depends on the
approximations of \eq{\ref{eq:discretisation}}. Expanding $f(\bfx)$ to the first order, gives
\begin{equation}
  \int f(\bfx) \psi_i(\bfx)\,d\bfx = f_i \int \psi_i(\bfx)\,d\bfx + (\nabla f)_i\cdot \int
  (\bfx-\bfx_i) \psi_i(\bfx)\,d\bfx.
\end{equation}
The first term is $f_i V_i$, and we can rewrite the integral in the second term in the
following form
\begin{equation}
  \int (\bfx-\bfx_i) \psi_i(\bfx)\,d\bfx = \int \bfy w(\bfx_i + \bfy) W(\bfy)\, d\bfy,
\end{equation}
where in the right hand side we changed variables from $\bfx$ to $\bfy = \bfx -
\bfx_i$. If we require that $w(\bfx_i + \bfy_j) d\bfy_j \approx C_i$ is approximately
constant in the neighbourhood of an $i$-particle, we can discretise the integral on the
right hand side to obtain $C_i \sum_j \bfy_j W(\bfy_j)$. Hence, if the particles are relaxed
such that this sum is minimised, the start up noise becomes negligible. Empirically, we
found that the particle distribution is regularised if the following quantity is minimised
\begin{equation}
  \delta{\cal R} = \sum_i |\Delta {\bf R}_i|^2,
  \label{eq:regular_min1}
\end{equation}
where 
\begin{equation}
  \Delta {\bf R}_i = \sum_j ({\bf x}_j - {\bf x}_i)W({\bf x}_j - {\bf x}_i, h_i),
  \label{eq:regular_min2}
\end{equation}
Ideally, $\delta{\cal R}$ should be equal to zero, but this is appears to be only possible
if particles are arranged on a lattice, for example a cubic or hexagonal close-packed
lattice. If particles are sampled randomly, which is more desirable in many problems,
their positions must be adjusted until \eq{\ref{eq:regular_min1}} reaches its
(approximate) minimum before assigning fluid state to the particles; this will reduce the
start-up noise in a simulation. Afterwards, this particle distribution can be used to
assign initial conditions for a problem at hand. To regularise, or to relax, particle
distribution, we use the following iterative procedure. First, we compute
\eq{\ref{eq:regular_min2}} for all particles. Afterwards, $i$-particle position is
updated: ${\bf r}_i^{n+1} = {\bf r}_i^{n} -\alpha \Delta{\bf R}_i$, where $\alpha < 0.1$;
such update reduces $\delta{\cal R}$. This operation is repeated until $\delta{\cal R}$ is
reached its minimum, or a desired minimal value.

\subsection{Time marching}\label{sect:time_marching}
Due to Godunov's nature of the particle conservation laws, it is tempting to employ
Hancock scheme (e.g. \citealp{citeulike:3115494}) to achieve a single-stage second-order
accurate time integration. However, such scheme does not include a predictor for
transverse waves in multi-dimensional Riemann problem, and therefore these remain only
first-order accurate. In multiple dimensions, unsplit numerical schemes usually adopt
Corner-Transport-Upwind method (CTU, \citealp{1990JCoPh..87..171C, 2008ApJS..178..137S})
for one-step second-order time integration, but applicability of CTU to meshless schemes
is not clear. Nevertheless, higher than the first order accurate time integration can be
achieved with multi-stage total-variation diminishing (TVD) Runge-Kutta methods
\citep{Gottlieb98totalvariation}. Here, we use a two-stage second order TVD Runge-Kutta
time marching scheme
\begin{equation}
  (V{\bf u})_i^{\rm p} = (V{\bf u})_i^0 + \dot{(V{\bf u})}_i^0 \Delta t,
\end{equation}
\vspace{-0.5cm}
\begin{equation}
  (V{\bf u})_i^1 = \frac{1}{2}\left[(V{\bf u})_i^{\rm p} + (V{\bf u})_i^0 + \dot{(V{\bf
      u})}_i^{\rm p} \Delta t \right],
\end{equation}
where $\dot{(V {\bf u})}_i^0$ and $\dot{(V {\bf u})}_i^{\rm p}$ are time derivatives of an
$i$-particle computed from $(V{\bf u})_i^0$ and $(V{\bf u})_i^{\rm p}$ respectively via
\eq{\ref{eq:meshless_1D}}, $V_i$ is the effective volume of $i$-particle, and $\Delta t =
   {\tt cfl}\times\min(L_i/c_{i, {\rm sig}})$, where $L_i = V_i$, $\sqrt{2 V_i/\pi}$ and
   $(3V_i/4\pi)^{1/3}$ for 1D, 2D and 3D respectively is a measure of particle linear
   size, $c_{\rm sig}$ is particle's signal speed (speed of sound for HD and of the fast
   magnetosonic wave for MHD), and ${\tt cfl} < 1$ is a usual Courant-Fridrisch-Levy
   number.

Particle positions obey the following equation of motion
\begin{equation}
  \frac{d{\bf x}_i}{dt} = {\bf a}_i,
\end{equation}
where ${\bf a}_i$ is particle's velocity. In practice we set it equal to the fluid
velocity, ${\bf a}_i = {\bf v}_i$, and the time integration is carried out in
drift-kick-drift approach. Namely, the particles are first drifted from their current
positions, $\bfx^0$, to the position at half time-step
\begin{equation}
  {\bf x}_i^{\rm h} = {\bf x}_i^0 + \frac{1}{2} {\bf a}_i^0 \Delta t.
\end{equation}
This particle distribution is used to compute smoothing lengths via
\eq{\ref{eq:constraint_h}} and other geometric quantities. Afterwards, we apply a two
stage TVD Runge-Kutta method to perform an update from $(V{\bf u})_i^0$ to $(V{\bf
  u})_i^1$ while keeping particles fixed in space and setting $V_i^0 = V_i^{\rm p} =
V_i^{\rm h}$, where $V_i^{\rm h}$ is $i$-particle volume at half time-step. Finally, the
particles are drifted for another half time-step with the updated fluid velocity, ${\bf
  a}^1 = {\bf v}^1$,
\begin{equation}
  {\bf x}_i^1= {\bf x}_i^{\rm h} + \frac{1}{2} {\bf a}_i^1 \Delta t.
\end{equation}

\subsection{Non-conservative formulation}
In the case of an HD or MHD system, the conservative formulation updates total energy
instead of thermal; thermal energy is obtained by subtracting magnetic and kinetic
energies from the total energy. When the supersonic advection is present, the sum of
thermal and magnetic energies, $U = E_{\rm th} + E_{\rm mag}$, is obtained by subtracting
two large numbers, namely total energy and kinetic energy. To avoid this, we suggest an
alternative non-conservative formulation, which evolves $U$ instead of the total
energy. Writing $E = U + {\bf P}^2/2M$, where ${\bf P}$ and $M$ are momentum and mass
respectively, gives
\begin{equation}
  \frac{dU}{dt} = \frac{dE}{dt} - \frac{d}{dt}\left(\frac{{\bf P}^2}{2M}\right).
\end{equation}
The latter term can be rewritten as ${\bf v}\cdot\dot{\bf P} - {\bf v}^2\dot M/2$
resulting in the following equations for $U$
\begin{equation}
  \frac{dU}{dt} = \frac{dE}{dt} - {\bf v}\cdot\frac{d{\bf P}}{dt} + \frac{{\bf
      v}^2}{2}\frac{dM}{dt}.
  \label{eq:nonconserv}
\end{equation}
Here, $d{\bf P}/dt$ and $dM/dt$ are time derivatives computed from conservative meshless
equations. In this form, the total energy will not be conserved to the machine accuracy,
but rather to the truncation error of the time-marching scheme. In other words, the
integration error is now lost from the system instead of appearing in the thermal
energy. Furthermore, the total energy can now be used as a quality control indicator of a
simulation.

\subsection{Modification of existing SPH codes}
The existing SPH codes which use conservative SPH formulation \citep{2002MNRAS.335..843M,
  2002MNRAS.333..649S} can be straightforwardly modified to implement our weighted
particle scheme due to similarity of SPH equations of motions and our equations of
meshless conservation laws, \eq{\ref{eq:meshless_1D}}. In particular, the neighbour search
should be modified such that \eq{\ref{eq:constraint_h}} is solved, which constrains the
number of particles, instead of the enclosed mass, in the neighbour sphere. At the end of
this step, the $h_i$ for each particle will be known that permits to compute volume of a
particle, $V_i = w_i(h_i)$. This volume is required to convert conservative fluid
variables, $(V {\bf u})_i = V_i{\bf u}_i$, to primitive ones ${\bf w}_i$. Afterwards, the
first loop is carried out over gather neighbours which computes the renormalisation
matrix, \eq{\ref{eq:renorm_matrix}}, and gradients of primitive fluid variables. The
second neighbour loop is required to compute the limiting functions,
\eq{\ref{eq:limiter}}, and this loop must be carried out over both gather and scatter
neighbours due to need to limit reconstruction to each of the interacting
particles. Finally, in the third neighbour loop the interactions between the particles are
computed. This is done in exactly the same way as in SPH, except that for every
$j$-neighbour of an $i$-particle, the fluid states are reconstructed at the midpoint,
$\bfx_{ij} = (\bfx_{i} + \bfx_{j})/2$ through
\eq{\ref{eq:linear_reconstruction}}. Finally, these together with the vector ${\bf
  n}_{ij}$ in \eq{\ref{eq:meshless_1D}} and the interface velocity ${\bf a}_{ij} = ({\bf
  v}_i + {\bf v}_j)/2$, are used in the Riemann solver to compute the interface fluxes,
$(\bar{\bf F}_{ij}\cdot \hat{\bf n}_{ij})$, which we describe in the following
sections. At the end of this final neighbour loop, one will have time derivative that
should be used to update the {\emph conservative} fluid variables. The global conservation
of mass and other conservative quantities, in the absence of source terms, is maintained
to the machine precision {\it independently} of the particle distribution. This also
includes total energy, unless \eq{\ref{eq:nonconserv}} is used, in which case the total
energy is conserved to the truncation error of a time integration scheme.

\section{Fluid dynamics}
Our meshless conservative equations can be applied to a system which can be written in the
form of conservation laws. The physics in this case is completely described by the source
terms, ${\cal S}$, and fluxes, ${\cal F}$. The latter can be obtained by solving an
appropriate one-dimensional Riemann problem between a particle and its neighbours. Hence,
the \eq{\ref{eq:meshless_1D}} can be applied to a variety of problems, such as
hydrodynamics, magnetohydrodynamics, and radiative transfer in the flux-limited diffusion
approximation. In what follows, we present application of our scheme to both ideal
hydrodynamics and magnetohydrodynamics.

\subsection{Ideal hydrodynamics}\label{sect:ideal_hd}
The Euler equations of ideal hydrodynamics in a frame moving with the velocity ${\bf a}$
read
\begin{equation}
  \frac{\partial{\cal U}}{\partial t} + \nabla\cdot({{\cal F} - {\bf a}\,{\cal U}}) = {\cal S},
\end{equation}
where 
\begin{equation}
  {\cal U} = \left(
  \begin{array}{c}
    \rho \\ 
    e_{\rm tot} \\
    \rho {\bf v} \\
  \end{array}
  \right), \quad
  {\cal F} = \left(
  \begin{array}{c}
    \rho{\bf v} \\ 
    (e_{\rm tot} + p){\bf v}\\
    \rho {\bf v}\otimes{\bf v} + p {\cal I} \\
  \end{array}
  \right),\quad
  {\cal S} = \left(
  \begin{array}{c}
    0 \\
    0 \\
    {\bf 0} \\
  \end{array}
  \right),
\end{equation}
where ${\cal I}$ is a unit tensor, and other symbols have their usual meaning. 

To obtain fluxes, an 1D Riemann problem is solved between two particles in the ${\bf
  n}_{ij}$ direction. This is accomplished by defining the rotation matrix, ${\cal A}$,
such that in the new coordinate system ${\bf n}_{ij}$ coincides with the $x'$-axis,
i.e. ${\bf n}'_{ij} = {\cal A}{\bf n}_{ij} = (|{\bf n}_{ij}|,0,0)$.  This transformation
is applied to all vector quantities from both the left and the right states, the scalar
quantities are left untouched. For example, the velocity transformation results in ${\bf
  v}'_{K} = {\cal A}{\bf v}_{K} = (v'_{x,K}, v'_{y,K},v'_{z,K})$, where $K=L$ for the left
and $K=R$ for the right state. These transformed states are the initial conditions of the
Riemann problem for 1D Euler equations
\begin{equation}
  \frac{\partial {\cal U'}_{1D}}{\partial t} + 
  \frac{\partial {\cal G'}_{1D}}{\partial x} =  0,
  \label{eq:hydro_1D}
\end{equation}
where, ${\cal G'}_{1D} = {\cal F'}_{1D} - a'_x {\cal U'}_{1D}$ and 
\begin{equation}
  {\cal U'}_{1D} = \left(
  \begin{array}{c}
    \rho \\ 
    e_{\rm tot} \\
    \rho v'_x\\
    \rho v'_y\\
    \rho v'_z
  \end{array}
  \right), \quad
  {\cal F'}_{1D} = \left(
  \begin{array}{c}
    \rho v'_x \\ 
    (e_{\rm tot} + p) v'_x\\
    \rho v'_x v'_x + p  \\
    \rho v'_y v'_x \\
    \rho v'_z v'_x 
  \end{array}
  \right).
\end{equation}
The space discretisation of this equation has the following form
\begin{equation}
  \frac{d{\cal U'}_i}{dt} + \frac{{\cal G'}_{i+\frac{1}{2}} - {\cal
      G'}_{i-\frac{1}{2}}}{\Delta x} = 0,
\end{equation}
where $\Delta x$ is size of the grid cell in 1D, and ${\cal G}'_{i+1/2}$ is an interface
flux between cells $i$ and $i+1$. This is the flux required to substitute into
\eq{\ref{eq:meshless_1D}}, and it can be obtained from the solution of an appropriate 1D
Riemann problem.

The solution of this 1D Riemann problem gives the fluxes, ${\cal G'}$, for each of the
component of ${\cal U'}$ in direction ${\bf n}_{ij}$. While the fluxes of the scalar
components can be directly used in \eq{\ref{eq:meshless_1D}}, those of a spatial vector,
however, must be rotated back to the original coordinate system because the flux of an
$x$-component of the vector, rather than $x'$, is required in the direction ${\bf
  n}_{ij}$. In the case of the velocity vectory, one first combines $(v'_x, v'_y,
v'_z)$-flux into a vector ${\bf G'}({\bf v}')=({\cal G'}(v'_x), {\cal G'}(v'_y), {\cal
  G'}(v'_z))$. Then an inverse transformation is applied to compute $(v_x, v_y,
v_z)$-flux, namely ${\bf G'}({\bf v}) = {\cal A}^{-1}{\bf G'}({\bf{v'}}) = ({\cal
  G'}(v_x), {\cal G'}(v_y), {\cal G'}(v_z))$, and these fluxes are then substituted into
\eq{\ref{eq:meshless_1D}}.

This approach permits the use of Riemann solvers that directly approximate interface flux,
rather than fluid states. In what follows, the HLLC Riemann solver is used to compute the
flux. The HLLC solver requires only velocity estimates for certain characteristic waves,
and is oblivious to the exact form of the equation of state. As a result it can be easily
extended to equations of states other than that of ideal gas. Here, we present formulae of
the HLLC Riemann solver in moving frame which can be directly used in
\eq{\ref{eq:meshless_1D}}; for the detailed derivation we refer the interested reader to
published literature (e.g. \citealp{toro99} (\S10), \citealp{2005JCoPh.208..315M}).

\begin{figure}
  \center
  \includegraphics[scale=0.3]{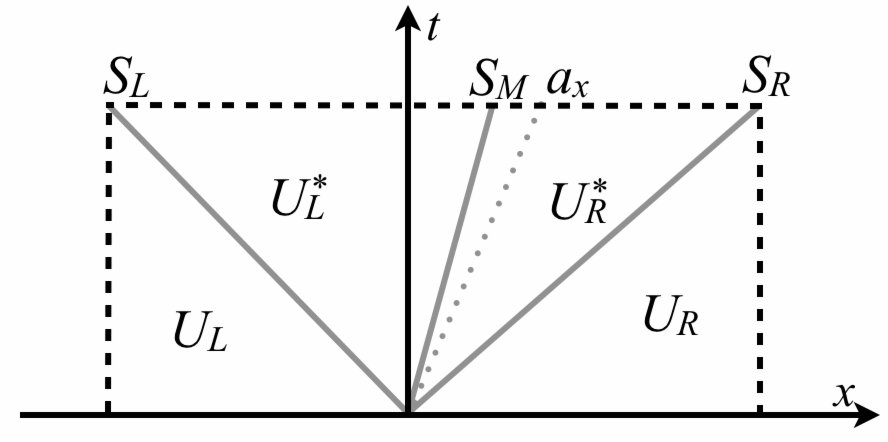}
  \caption{This figure shows the wave-structure of HLLC Riemann solver. Namely, HLLC
    solver resolves all three HD waves: left, right, and middle wave also known as contact
    discontinuity waves. The left and right wave can be either shock or rarefaction wave,
    or both. This solver approximates the structure of the two intermediate states,
    between $S_L$ and $S_M$, and $S_M$ and $S_R$, by a constant state.}
  \label{fig:HLLC}
\end{figure}

A typical wave-structure of HLLC solver is shown in \fig{\ref{fig:HLLC}}. The
velocity of left, right and middle waves are $S_L$, $S_R$ and $S_M$ respectively, and the
constant states sandwiched between these waves are $U'^*_L$ and $U'^*_R$ respectively. The
HLLC-flux at the interface moving with velocity $a'_x$ is
\begin{equation}
  {\cal G}'^{HLLC} = \left\{ 
  \begin{array}{ll}
    {\cal F}'_L - a'_x {\cal U}'_L                          & a'_x < S_L, \\
    {\cal F}'_L + S_L({\cal U}'^*_L - {\cal U}'_L) - a'_x {\cal U}'^*_L   & S_L \leq a'_x \leq S_M, \\
    {\cal F}'_R + S_R({\cal U}'^*_R - {\cal U}'_R) - a'_x {\cal U}'^*_R   & S_M \leq a'_x \leq S_R, \\
    {\cal F}'_R - a'_x {\cal U}'_R                          & S_R \leq a'_x,
  \end{array}
  \right.
  \label{eq:flux_hllc}
\end{equation}
where ${\cal F}'_L = {\cal F}({\cal U}'_L)$ and ${\cal F}'_R = {\cal F}({\cal U}'_R)$. The intermediate states are defined by
\begin{equation}
  {\cal U}'^*_K = \rho_K \left(\frac{S_K - v'_{xK}}{S_K - S_M}\right)
  \left[
  \begin{array}{c}
    1    \\
    S_M   \\  
    v'_{yK} \\ 
    v'_{zK}  \\
    \frac{E_K}{\rho_K} + (S_M - v'_{xK})\left(S_M + \frac{p_K}{\rho_K(S_K - v'_{xK})}\right)
  \end{array}
  \right],
\end{equation}
for $K = L$ and $K = R$. Finally, estimates of wave speeds are $S_L = \min(v'_{xL},
v'_{xR}) - c_{\rm s}$ and $S_R = \max(v'_{xL}, v'_{xR}) + c_{\rm s}$, where $c_{\rm s}
= \max(c_{{\rm s} L}, c_{{\rm s} R})$. Other wave-speed estimates, such as those based
on Roe-averages, can be used as well. Finally, the speed of the middle wave and the
pressure in the $\star$-states is given by the following formulae
\begin{equation}
  S_M = \frac{p_R - p_L + \rho v'_{xL} (S_L - v'_{xL}) - \rho v'_{xR}(S_R - v'_{xR})}
  {\rho_L(S_L - v'_{xL}) - \rho_R(S_R - v'_{xR})},
\end{equation}
\begin{equation}
  p^\star = \frac{(S_R - v_{xR})\rho_R p_L - (S_L - v_{xL})\rho_L p_R +
    \rho_L\rho_R(S_R - v_{xR})(S_L - v_{xL})(v_{xR} - v_{xL})}{(S_R - u_R)\rho_R - (S_L - u_L)\rho_L},
\end{equation}
with $v^\star_{xL} = v^\star_{xR} = S_M$ and $p^\star_L = p^\star_R = p^\star$.

\subsection{Ideal MHD}\label{sect:ideal_mhd}

Our meshless conservative equations can also be applied to the equations of ideal MHD. In
particular, we can rewrite MHD equations in the conservative form with the following
conservative variables, fluxes and source terms
\begin{equation}
  {\cal U} = \left(
  \begin{array}{c}
    \rho \\ 
    e_{\rm tot} \\
    \rho {\bf v} \\
    {\bf B} \\
  \end{array}
  \right), \quad
  {\cal F} = \left(
  \begin{array}{c}
    \rho{\bf v} \\ 
    (e_{\rm tot} + P_T){\bf v} - ({\bf v}\cdot{\bf B}) {\bf B}\\
    \rho {\bf v}\otimes{\bf v} + P_T{\bf I} - {\bf B}\otimes{\bf B}\\
    {\bf v}\otimes{\bf B} - {\bf B}\otimes{\bf v} \\
  \end{array}
  \right),\quad
  {\cal S} = \left(
  \begin{array}{c}
    0 \\
    0 \\
    {\bf 0} \\
    {\bf 0} \\
  \end{array}
  \right),
\end{equation}
here $P_T = p + {\bf B}^2/2$ is sum of the thermal and magnetic pressures. The biggest
difficulty in solving these equations is to maintain $\nabla\cdot{\bf B} = 0$
constraint. While there are several ways to satisfy this constraint in finite-difference
methods, it is not clear how this can be done in meshless schemes. In SPH,
\cite{2007MNRAS.379..915R} were able to circumvent this problem via use of the Euler
potentials, $\alpha$ and $\beta$, which in smooth flows are advected with the flow, and
are used to compute magnetic field ${\bf B} = \nabla\alpha \times \nabla\beta$. While
mathematically this guarantees zero-divergence, in practice, however, the divergence is
non-zero because of non-commuting nature of SPH cross-derivatives. This is also holds for
the vector potential, ${\bf A}$, which may explain unstable behaviour of SPMHD equations
with vector potential \citep{2010MNRAS.401.1475P}. While SPH equations appear to be stable
in the Euler potential formulation, they are unable to model topologically non-trivial
field configurations \citep{2010MNRAS.401..347B}, and therefore are not suited to simulate
complex magnetic field evolution, such as winding of magnetic field lines and
magneto-rotational instability.  Motivated by this, and by the fact that the existing MHD
Riemann solvers use magnetic field as a primary quantity, we chose to evolve ${\bf B}$
instead, and apply one of the divergence cleaning methods known to work in Godunov
finite-difference MHD schemes.

In their paper, \cite{1999JCoPh.154..284P} showed that a self-consistent MHD equations
must include source terms proportional to $\nabla\cdot{\bf B}$ to insure stability and
Galilean invariance
\begin{equation}
  {\cal S} = -\nabla\cdot{\bf B}\left(
  \begin{array}{c}
    0 \\
    {\bf v}\cdot {\bf B} \\
    {\bf B} \\
    {\bf v} \\
  \end{array}
  \right).
\end{equation}
While this formulation both removes the instabilities associated with non-zero divergence
and tends to keep the divergence at the truncation level, the divergence can still grow in
certain situations. To avoid this, we also include a Galilean invariant form of the
hyperbolic-parabolic divergence cleaning method due to \cite{2002JCoPh.175..645D}, which
results in the following system
\begin{equation}
  {\cal U} = \left(
  \begin{array}{c}
    \rho \\ 
    e_{\rm tot} \\
    \rho {\bf v} \\
    {\bf B} \\
    \rho \psi \\
  \end{array}
  \right), \quad
  {\cal F} = \left(
  \begin{array}{c}
    \rho{\bf v} \\ 
    (e_{\rm tot} + P_T){\bf v} - ({\bf v}\cdot{\bf B}) {\bf B}\\
    \rho {\bf v}\otimes{\bf v} + P_T{\bf I} - {\bf B}\otimes{\bf B}\\
    {\bf v}\otimes{\bf B} - {\bf B}\otimes{\bf v} \\
    \rho\psi {\bf v}\\
  \end{array}
  \right),\quad
  {\cal S} = \left(
  \begin{array}{l}
    \qquad0 \\
    -(\nabla\cdot{\bf B}) {\bf v}\cdot{\bf B} - {\bf B}\cdot\nabla\psi \\
    -(\nabla\cdot{\bf B})\,{\bf B} \\
    -(\nabla\cdot{\bf B})\,{\bf v} - \nabla\psi \\
    -(\nabla\cdot{\bf B}) c_h^2\rho - \psi\rho/\tau \\
  \end{array}
  \right).
\label{eq:8wave_mhd}
\end{equation}
Here, $c_h$ and $\tau$ is the speed and the damping timescale of the divergence wave, also
known as 8-wave, respectively. Usually, $c_h$ is set to be highest characteristic speed,
i.e. that of the fast magnetosonic wave, $\tau = L/(c_{\bf r } c_h)$, where $L$ is an
effective size of the particle defined in \textsection\ref{sect:time_marching}, and
$c_{\bf r}$ is a constant, which, following \cite{2010JCoPh.229.2117M}, we set $c_{\bf r}
= 0.03$.

In their paper, \cite{2002JCoPh.175..645D} also presented a method to generalise an
arbitrary 1D MHD Riemann solver to include an additional scalar field $\psi$ which
transports the divergence away form the source and damps it. In general, the left and
right reconstructed states in coordinate system ${\cal A}$ (see \S\ref{sect:ideal_hd})
have discontinuous $B'_x$--the magnetic field component normal to the interface. The 1D
Riemann solvers, however, require this field to be continuous, and this can be computed
with the following equations
\begin{equation}
  \bar B'_x = \frac{1}{2}(B'_{xL} + B'_{xR}) - \frac{1}{2 c_{h,ij}}(\psi_R - \psi_L),
  \label{eq:Bx_bar}
\end{equation}
\vspace{-0.5cm}
\begin{equation}
  \bar \psi = \frac{1}{2}(\psi_L + \psi_R) - \frac{c_{h,ij}}{2}(B'_{xR} - B'_{xL}).
  \label{eq:psi_bar}
\end{equation}
Here, $B'_{xK}$ are the left ($K= L$) and right ($K= R$) reconstructed states for $B'_x$
and $\psi$, and $c_{h,ij} = \max(c_{h,i}, c_{h,j})$. These interface values of $\bar B'_x$
and $\bar\psi$ are used to compute $\nabla\cdot{\bf B}$ and $\nabla\psi$ respectively,
which are required for the source terms in \eq{\ref{eq:8wave_mhd}},
\begin{equation}
  V_i (\nabla\cdot{\bf B})_i = -\sum_j \bar B'_x |{\bf n}_{ij}|,
  \label{eq:divB}
\end{equation}
\vspace{-0.5cm}
\begin{equation}
  V_i (\nabla\psi)_i = -\sum_j \bar\psi\,{\bf n}_{ij}.
\end{equation}
In Appendix \ref{sect:appendix}, we provide standard formulae to compute HLL and HLLD
fluxes for the 1D MHD Riemann problem, which take $\bar B'_x$ as the continuous normal
component of the magnetic field. Finally, the advection flux for the scalar $\rho\psi$ is
\begin{equation}
  F_\psi = F_\rho  \times \left\{
    \begin{array}{cc}
      \psi_L, & {\rm if} \,\, F_\rho > 0,  \\
      \psi_R, & {\rm otherwise},
    \end{array}
    \right. 
\end{equation}
where, $F_\rho$ is a mass flux given by the Riemann solver. This completes the description
of our meshless MHD scheme.

\section{Scheme validation}\label{sect:results}
The weighted particle scheme is validated on several standard problems for ideal
hydrodynamics and MHD problems, which are usually used to test hydrodynamic and MHD
schemes (e.g. \citealp{2000JCoPh.161..605T}, \citealp{2008ApJS..178..137S}.
In all simulations what follows, we use $N_{\rm ngb} = 19$ and $32$ in
\eq{\ref{eq:constraint_h}} for 2D and 3D simulations respectively. This choice is
motivated by the analogy with finite-difference schemes. In one dimension, the number of
the neighbouring cells that a given cell interacts with depends on the order of the
numerical scheme, and is usually two for the second order scheme. In SPH, two to four
neighbouring particles are usually used in one-dimensional simulations. If this number is
scaled to three dimensions, and taking into account that a kernel has spherical shape, the
estimated number of neighbours are 16 and 32 for 2D and 3D respectively. The MHD
simulations in 2D appear noisy with $N_{\rm ngb} = 16$, which motivated us to use larger
$N_{\rm ngb}$. In principle, large $N_{\rm ngb}$ can be used, should this be necessary,
but this will decrease the resolution due to larger smoothing length.

\subsection{Shock tubes}


\subsubsection{Brio-Wu shock tube}
\begin{figure}
  \center \includegraphics[scale=0.3]{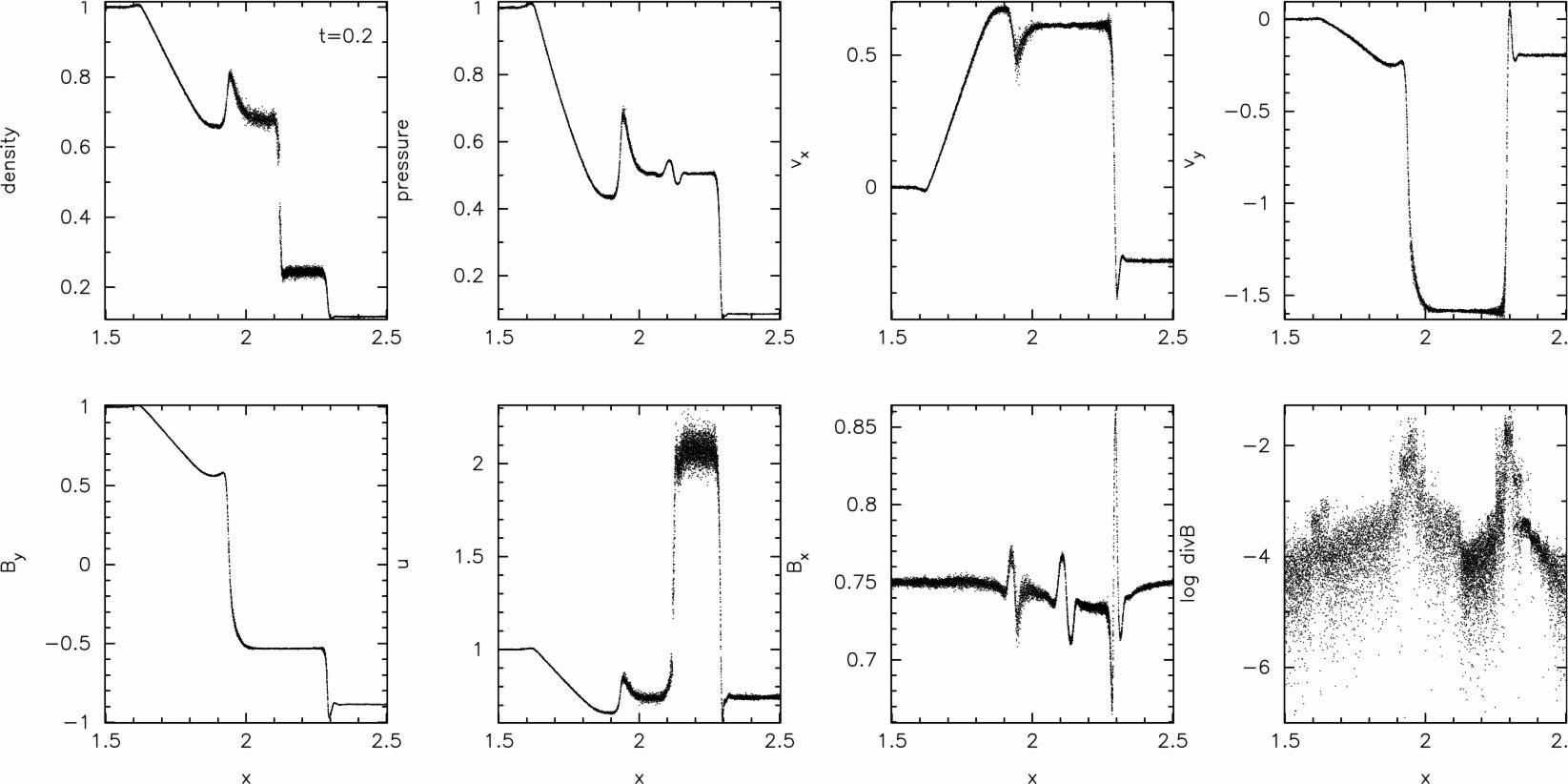}
  \caption{Solution to the Brio-Wu MHD shock tube problem at time $t = 0.2$. The panels
    display (from left to right, top to bottom) density, gas pressure, $x$- and $y$-
    velocity profiles, $B_y$, thermal energy $u = P/\rho$, $B_x$ and the logarithm of
    ${\tt divB} = L\nabla\cdot{\bf B}/|{\bf B}|$, where $L$ is an effective size of a
    particle (\S\ref{sect:time_marching}).}
  \label{fig:brio_wu}
\end{figure}

This problem was introduced by \cite{1988JCoPh..75..400B} to test the ability of an MHD
scheme to accurately model shock waves, contact discontinuities and compound structures of
MHD. Here, the problem is solved in a periodic 2D domain of size $[0,4]\times[0,0.25]$
with randomly sampled $5\cdot10^4$ particles; this results in an effective resolution of
$895\times56$. The particles are initially relaxed before the initial conditions are
set. The left state, $x < 2$, is set with the following values: $\rho_L = 1$, $p_L = 1$,
$B_{yL} = 1$. The right state has $\rho_R = 0.125$, $p_R = 0.1$ and $B_{yL} = -1$. Both
states have zero initial velocities, $B_z = 0$ and $B_x = 0.75$. The problem is solved
with an ideal gas equation of states and $\gamma = 2$. Various profiles, e.g. density and
pressure, at time $t = 0.2$ are show in \fig{\ref{fig:brio_wu}}.

The results are overall consistent with 1D Godunov-MHD scheme, namely jumps across the
discontinuities and location of discontinuities. The two bottom right panels show the
value of $B_x$ field and the ${\tt divB} = L\nabla\cdot{\bf B}/|{\bf B}|$ as a function of
particle $x$-coordinate. The parallel magnetic field slightly deviates from its constant
value, except near discontinuities where it exhibits jumps. The divergence, however,
remains small, even across discontinuities. The existence of blip in pressure at location
of contact discontinuity, $x \approx 2.1$, and shock waves, $x \approx 2.3$ has the same
origin as in SPH: the particle distribution across a discontinuity is less regular in a
sense that the approximation in \eq{(\ref{eq:discretisation}} is not sufficient to provide
accurate results. This is a known issue in Godunov SPH, and higher order approximations
are able to reduce the amplitude of the blip \citep{2002JCoPh.179..238I}. Overall, the
solution obtained by the meshless scheme is in a good agreement with high-resolution 1D
Euleriean schemes.

\subsubsection{T\'{o}th shock tube}
\begin{figure}
  \center \includegraphics[scale=0.3]{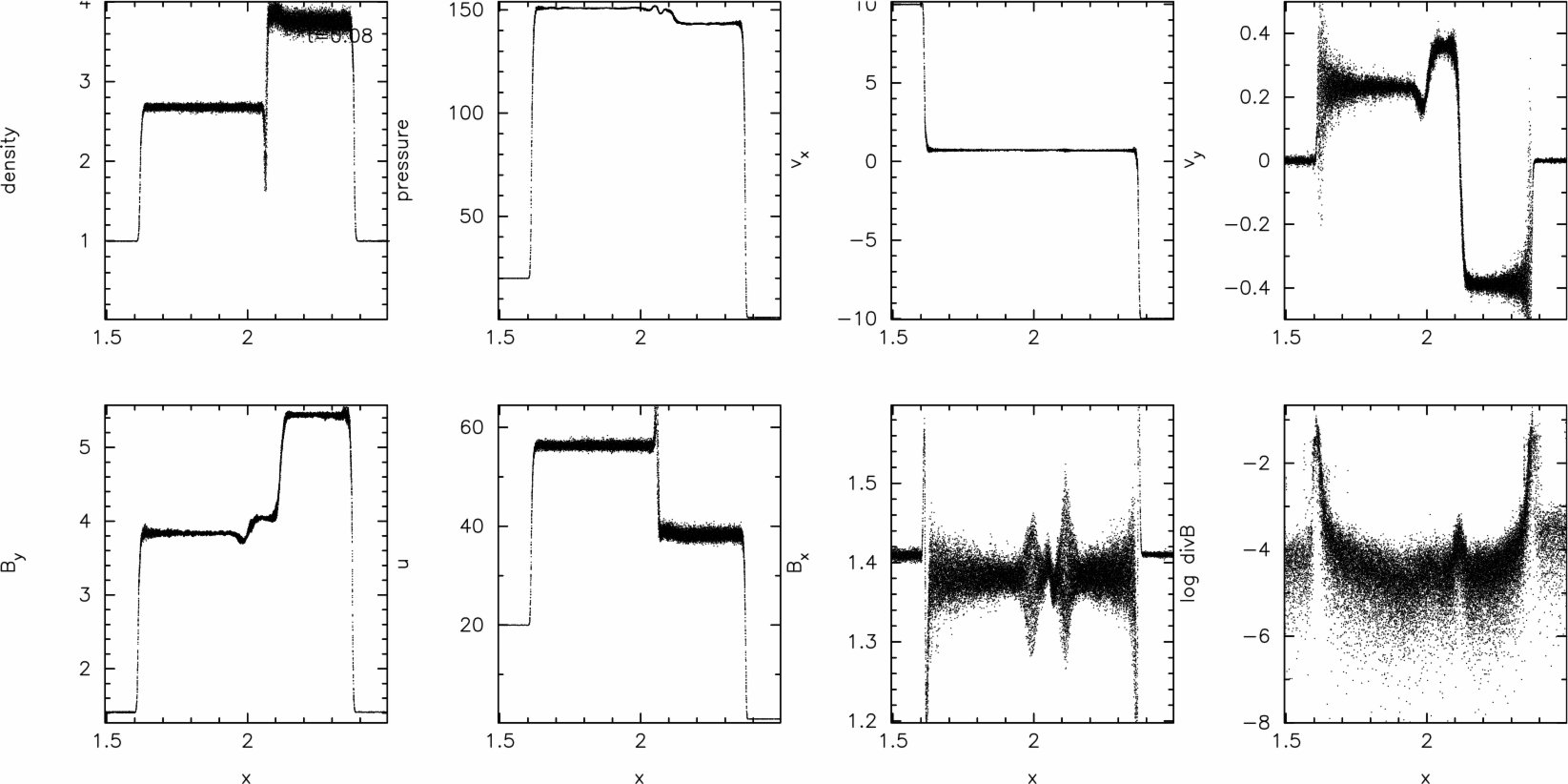}
  \caption{Solution to the T\'{o}th MHD shock tube problem at time $t = 0.2$. The panels
    display (from left to right, top to bottom) density, gas pressure, $x$- and $y$-
    velocity profiles, $B_y$, thermal energy $u = P/\rho$, $B_x$ and ${\tt divB} =
    L\nabla\cdot{\bf B}/|{\bf B}|$, where $L$ is an effective size of a particle.}
  \label{fig:toth_shock}
\end{figure}

Another challenging shock tube problem was introduced by \cite{2000JCoPh.161..605T}. In
this problem, two streams of magnetised gas supersonically collide with each
other. \cite{2000JCoPh.161..605T} showed that some MHD schemes with source terms
proportional to $\nabla\cdot{\bf B} = 0$ produce wrong jump conditions. This was
challenged by \cite{2010JCoPh.229.2117M}, who showed that if divergence is cleaned in
hyperbolic-parabolic manner \citep{2002JCoPh.175..645D}, the jump conditions are correct
even with the presence of the source terms. The problem is set in 2D periodic domain with
size $[0,4]\times[0,0.25]$ with randomly distributed $5\cdot 10^4$ particles. After the
particle distribution is relaxed, the following initial conditions are set. The left
states has $\rho_L = 1$, $p_L = 20$, $v_{xL} = 10$, and the right state has $\rho_R =
1.0$, $p_R = 1$, $v_{xR} = -10$. Both states have $B_{x} = B_{y} = 5/\sqrt{4\pi}$. This
problem is solved with an ideal gas equation of state and $\gamma = 5/3$.  The solution to
this problem is shown in \fig{\ref{fig:toth_shock}}, which can be compared to solutions
obtained by \citep{2000JCoPh.161..605T} and \cite{2010JCoPh.229.2117M}. Our meshless
scheme is able to recover correct jump conditions and to maintain constant $B_x$ field
within few percent accuracy, except across discontinuities. Furthermore, the divergence in
this problem remains small.

\subsection{Advection of a magnetic field loop}
This problem tests the ability of the scheme to transport magnetic loop across
computational domain. This problem is proven to be a stringent test for finite-difference
schemes. The computational domain in this test is a cuboid with dimensions
$[0,2]\times[0,1]\times[0,0.5]$ which is initially filled with $128\times64\times32$
particles on cubic lattice to assure zero noise. The fluid state is set to $\rho = 2$
inside the loop and $\rho = 1$ outside the loop, $p = 1$, ${\bf v} = (2, 1, 0.5)$ and
${\bf B} = (f(R) y, -f(R) x, 0)$, where $R = \sqrt{x^2 + y^2}$ and
\begin{equation}
  f(R) = \left\{
  \begin{array}{cl}
    B_0/R & 0 < R < R_0, \\
    0, & {\rm otherwise}.
  \end{array}
  \right.
\end{equation}
Here, $R_0 = 0.3$ is a radius of the loop, $B_0 = 10^{-3}$ is the initial magnetic field
strength that results in $2 \beta = P_{\rm gas}/P_{\rm mag} = 10^6$. With such high
$\beta$ magnetic field does not play dynamical role and should be transported as a passive
scalar. Periodic boundary conditions are used in this test, and the ideal gas equation of
state with $\gamma = 5/3$.

\begin{figure}
  \center
  \includegraphics[scale=0.4]{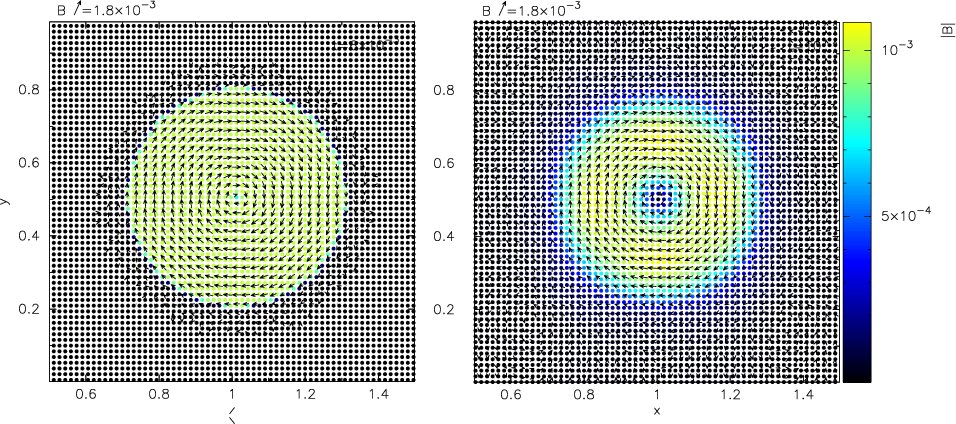}
  \caption{Amplitude (colours) and direction (vectors) of magnetic field at the beginning
    (left panel, $t = 0$) and at the end (right panel, $t = 10$) of the simulation. The
    magnetic field distribution is show in the $XY$-plane passing through $z = 0.25$. In
    the right panel, the strength of magnetic field in the centre is close to zero due to
    numerical resistivity, which is consistent with finite-difference calculations.}
  \label{fig:advection0}
\end{figure}

\begin{figure}
  \center
  \includegraphics[scale=1.0]{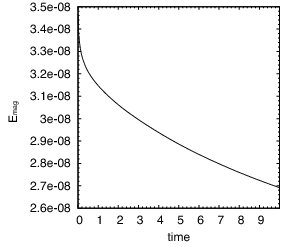}
  \caption{This figure demonstrates decay of magnetic energy during advection of magnetic
    field loop. The magnetic energy decay is due to numerical resistivity introduced by
    divergence cleaning procedure, and not through the solution of an MHD Riemann
    problem.}
  \label{fig:Emag}
\end{figure}

\begin{figure}
  \center
  \includegraphics[scale=0.4]{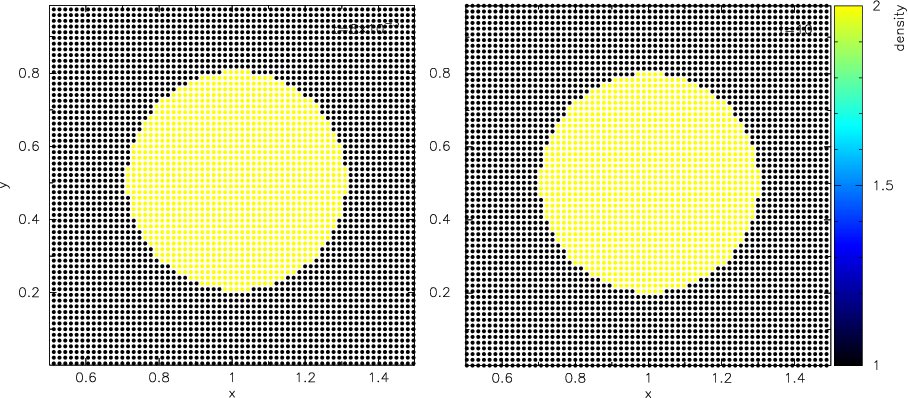}
  \caption{This figure shows that our scheme is able to transport mass, and therefore
    passive scalars, without any numerical diffusion in a constant velocity field. The
    right and left panels show plots of the density field in the $XY$-plan passing through
    $z = 0.25$, at the beginning and the end of the simulation respectively.}
  \label{fig:advection10}
\end{figure}

In \fig{\ref{fig:advection0}} we shows magnetic field structure at the $t = 0$ and $t =
10$, which corresponds to ten crossings of the computational domain, and in
\fig{\ref{fig:Emag}} we show the magnetic energy as a function of time. This figures
demonstrate the ability of the meshless scheme to advect magnetic loop quite
well. Furthermore, the decay of the magnetic energy is at least as slow as resulted from
high-order Godunov MHD schemes. This is certainly expected in light of semi-Lagrangian
nature of the scheme. One may however expect that the energy should not decay at all since
the magnetic field is transported as a passive scalar. Indeed, \fig{\ref{fig:advection10}}
demonstrates that the scheme is able to transport mass, and therefore passive scalar,
without any diffusion. The difference with magnetic field stems from the different nature
of equation that transport mass, or advect passive scalar, and the induction equation as
implemented in the scheme. In fact, the decay is caused solely by the diffusion of
magnetic field due to divergence cleaning equations, i.e.  \eq{(\ref{eq:Bx_bar}} and
\eq{\ref{eq:psi_bar}}. The fluxes resulting from HLLD Riemann solver are in fact zero
since both particles and the frame move with the same velocity. The divergence of the
magnetic field, even though is zero analytically, is not necessary zero in the
discretisation set by \eq{\ref{eq:Bx_bar}} and\eq{\ref{eq:divB}}, and this produces
evolution of magnetic field due to non zero value of $\nabla\psi$ in the source terms of
\eq{\ref{eq:8wave_mhd}}. Among all discretisations studied by \cite{2000JCoPh.161..605T},
this is the special one because of its use in the discretisation of Maxwell stress term to
obtain Lorentz force. If the $\nabla\cdot{\bf B} = 0$ in this discretisation, no force
parallel to magnetic field exists, and therefore the source terms proportional to this
divergence vanishes. Incidentally, this is the discretisation of divergence that is
enforced to zero in constrained transport (CT) formalism \citep{1988ApJ...332..659E}. More
importantly, the maintenance of zero divergence in other discretisations, such as
cell-centred, does not guarantee vanishing divergence in CT-discretisation, but this will
probably be small thus giving minimal damage to the solution.

\subsection{Blob test}
An interesting and challenging problem to test particle hydrodynamic schemes has been
proposed by \cite{2007MNRAS.380..963A}. This problem demonstrates the destructive property
of the Kelvin-Helmholtz instability (KHI) in three-dimensional simulations. The setup
consist of a dense cloud with $\rho_{\rm cl} = 10$ moving supersonically through a less
dense ambient fluid with $\rho_{\rm amb} = 1$. Initially, the cloud is at rest and in the
pressure equilibrium with the ambient fluid, at $p_{\rm amb} = 1$. The velocity of the
ambient medium is ${\bf v} = (2.7 c_{\rm s,amb}, 0, 0)$, where $c_{\rm s,amb}$ is its
sound speed. The cloud radius is $R_{\rm cl} = 0.1$, and an ideal gas equation of state is
used with $\gamma = 5/3$. The initial magnetic field is set to zero, and it will stay so
throughout the simulation. This problem is set in a periodic domain with dimensions
$[0,3]\times[0,1]\times[0,1]$, in which $7\cdot 10^5$ particles are sampled in the strip
$|y - 0.5| < 1.1 R_{\rm cl}$ and $3\cdot 10^5$ outside. This setup permits to resolve
cloud and the impacting ambient fluid with high resolution ($h \sim 0.01$) for a total of
$10^6$ particles. The particles are sampled from a three-dimensional Sobol quasi-random
sequence \citep{1992nrca.book.....P}. To remove the initial noise, this initial particle
distribution is relaxed before the initial values for the density field, pressure and
velocity are set.

\begin{figure}
  \center \includegraphics[scale=0.7]{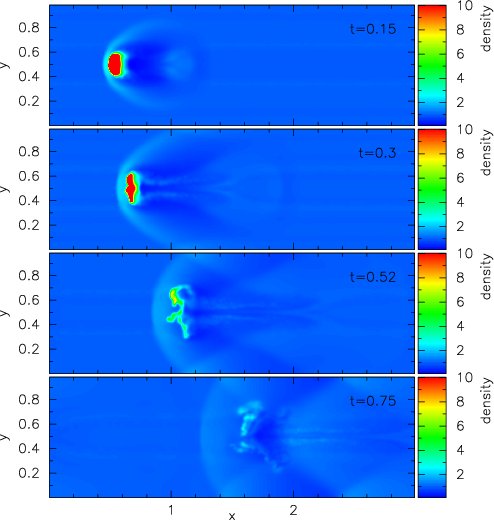}
  \caption{Density field plot in $XY$-plane passing through $z = 0.5$. The panels show
    density profile at time $0.5$, $1.0$, $1.5$ and $2.5$ $T_{\rm KH}$, where $T_{\rm KH}
    \approx 0.3$. At $t = 0.5 T_{\rm KH}$ (top panel), the cloud shape is distorted by RAM
    pressure. At $t = T_{\rm KH}$ (second panel from the top), the KHI instability deforms
    the shape of the cloud further. The cloud destruction begins at $t = 1.5 T_{\rm KH}$,
    and by $t = 2.5 T_{\rm KH}$ the cloud lost more than 90\% of its mass.}
  \label{fig:blob_dens}
\end{figure}

\begin{figure}
  \center
  \includegraphics[scale=1.0]{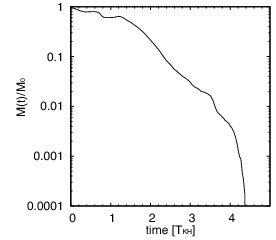}
  \caption{This plot show cloud mass as a function of time. In agreement with
    finite-difference scheme the cloud looses most of its mass at $t \approx 2$ $T_{\rm
      KH}$, and is completely destroyed at $t \gtrsim 3 T_{\rm KH}$.}
  \label{fig:blob_mass}
\end{figure}

Following \cite{2007MNRAS.380..963A}, the Kelvin-Helmholtz timescale is defined $T_{\rm
  KH} = 1.6 \tau_{\rm cr}$, where $\tau_{\rm cr} = 2R_{\rm cl} \sqrt{\rho_{\rm
    cl}/\rho_{\rm amb}}/v_{\rm amb}$. For parameters used in this simulations, this gives
$T_{\rm KH} \approx 0.3$. The snapshots of the density distribution in plane $z = 0.5$ at
$t = 0.5, 1.0, 1.5$ and $2.5\, T_{\rm KH}$ are shown in \fig{\ref{fig:blob_dens}}. These
are in an excellent agreement with those presented in \cite{2007MNRAS.380..963A}. The time
dependence of cloud mass is shown in \fig{\ref{fig:blob_mass}}, where a particle is
considered to be part of a cloud if $\rho > 0.64 \rho_{\rm cl}$ and $T < 0.9 T_{\rm
  amb}$. In agreement with finite-difference methods, the cloud lost nearly 90\% of its
mass within $2\,T_{\rm KH}$.

\subsection{Spherical blast-wave}\label{sect:blast}
\begin{figure}
  \center
  \includegraphics[scale=0.5]{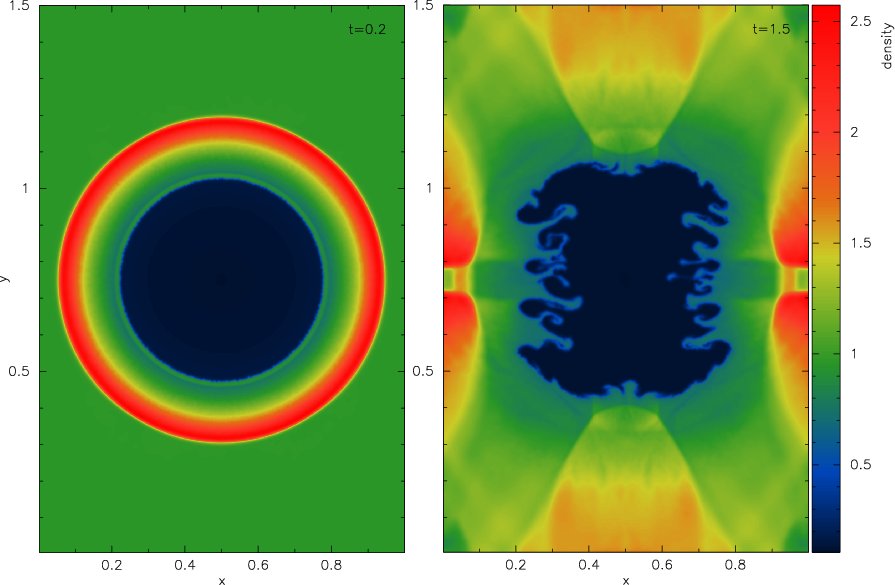}
  \caption{This figure show the density at $t = 0.2$ and $t = 1.5$ for non-magnetised
    spherical blast wave problem. The location of the shock front is in agreement with
    high-resolution conservative numerical schemes \citep{2008ApJS..178..137S}. Right
    panel show dense fingers in rarefied media which are formed by Richtmyer-Meshkov
    instability. This demonstrates that our scheme is able to capture important fluid
    instabilities without any fine-tuning.}
  \label{fig:blast_hd}
\end{figure}

\begin{figure}
  \center \includegraphics[scale=0.5]{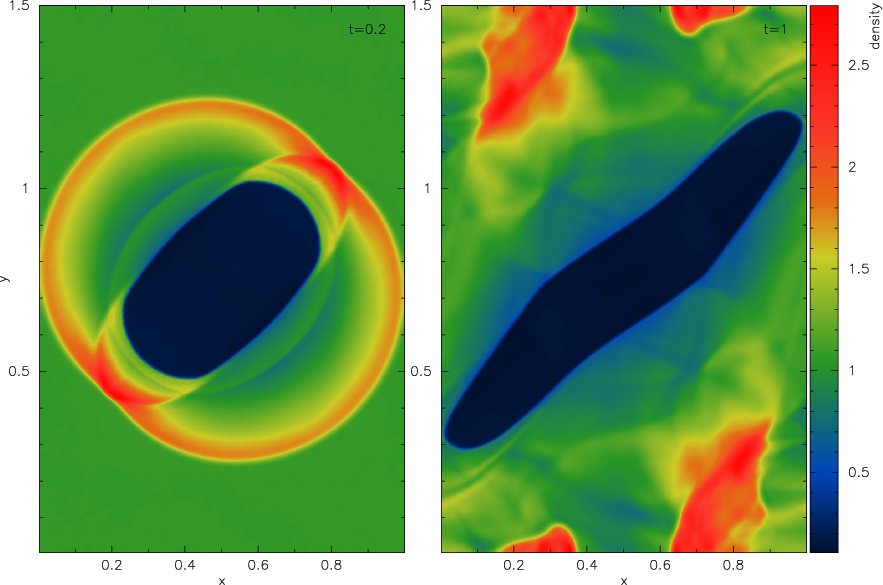}
  \caption{This figure show the density at $t = 0.2$ and $t = 1$ for magnetised spherical
    blast wave problem. The initial magnetic field in this problem has angle of $\pi/4$
    with $x$-axis. Since the shock moves more easily along the magnetic field lines, this
    explains the elongated shape of the shock-front, which contrasts with hydrodynamical
    case. As in hydrodynamical test, here the location of shock from is in excellent
    agreements with conservative Eulerian MHD schemes \citep{2008ApJS..178..137S}. The
    right panel, shows further evolution of the black wave, after shock reached the
    rarefied medium. In contrast to the hydrodynamical case, magnetic field inhibits
    development of Richtmyer-Meshkov instability.}
  \label{fig:blast_mhd}
\end{figure}

The problem is initiated with an overpressured central region in a uniform density and
magnetic field. The computational domain is a unit square filled with fluid with $\rho =
1$. Within $R < R_0 = 0.1$, the pressure is set to $10$, whereas outside $p = 0.1$. In
magnetised case, there is also a uniform magnetic field ${\bf B} = (1/\sqrt{2},1/\sqrt{2},
0)$. The equation of state is that of an ideal gas with $\gamma = 5/3$. The particles are
sampled in a periodic box $[0,1]\times[0,1.5]$, such that $5\cdot 10^4$ particles are
randomly sampled in three nested rectangles: $[0,1]\times[0.1,5]$, $[0.25,
  0.75]\times[0.375, 1.125]$ and $[0.375, 0.625]\times[0.5625, 0.9375]$. Before the
initial conditions are set, the particles are relaxed into a regular distribution to
reduce start-up noise. In figures \ref{fig:blast_hd} and \ref{fig:blast_mhd} we show
density plots for non-magnetised and magnetised cases respectively. Of particular interest
here is the ability of the particle weighted method to resolve Richtmyer-Meshkov
instability, shown in right panel of \fig{\ref{fig:blast_hd}}. In the magnetised case,
however, the presence of strong magnetic field inhibits development of this instability
(the right panel of \fig{\ref{fig:blast_mhd}}).

\subsection{Orszag-Tang vortex}\label{sect:orszag}
The Orszag-Tang vortex \citep{1979JFM....90..129O} is a standard test problem that is used
to validate many numerical MHD schemes. The setup involves periodic domain of size
$[0,1]\times[0,1]$ with an adiabatic equation of state with $\gamma = 5/3$. The initial
density and pressure are set in all computational domain to $25/(36\pi)$ and $5/(12\pi)$
respectively. The velocity ${\bf v} = (-\sin(2\pi y), +\sin(2\pi x), 0)$ and magnetic
field ${\bf B} = (-B_0\sin(2\pi y), B_0\sin(4\pi x), 0.0)$, where $B_0 =
1/\sqrt{4\pi}$. The second simulation involves the same initial conditions, with the
exception that the problem is solved a boosted frame, with the initial velocity ${\bf v} =
{\bf v}_{\rm rest} + {\bf v}_{\rm boost}$, where ${\bf v}_{\rm boost} = (10, 10, 10)$.
The simulation is solved with $10^5$ particles, where first $5\cdot 10^4$ where randomly
sampled within the square $[0,1]\times[0,1]$ and the second $5\cdot 10^4$ in the square of
half size, $[0.25, 0.75]\times[0.25,0.75]$. This permitted to achieve high resolution in
the central region of the problem. The initial conditions were set, as soon as
distribution was relaxed.

\begin{figure}
  \center
  \includegraphics[scale=0.5]{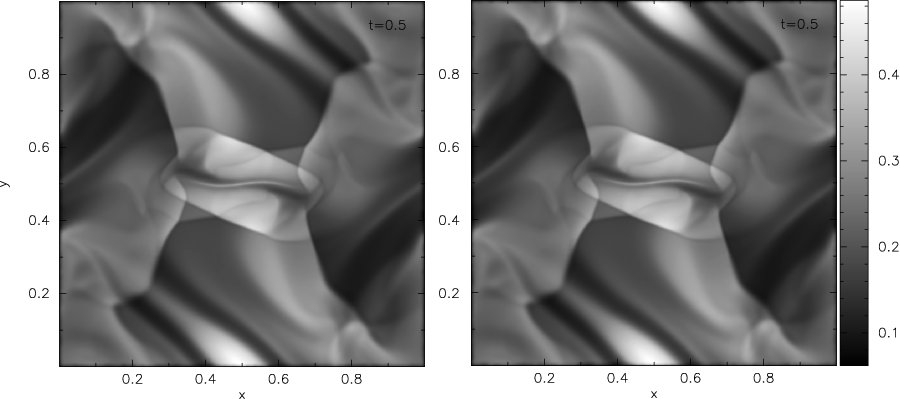}
  \caption{Density distribution of Orszag-Tang vortex at $t = 0.5$. The left and right
    panels show the result of the simulation in the rest and boosted frame
    respectively. We point out that even at low resolution our scheme is able to resolve
    the fine structure in the bottom left quadrant; in this quadrant, the effective
    resolution is $128\times 128$, compared to $256\times 256$ in the centre.}
  \label{fig:ot5_2d}
\end{figure}

\begin{figure}
  \center
  \includegraphics[scale=0.3]{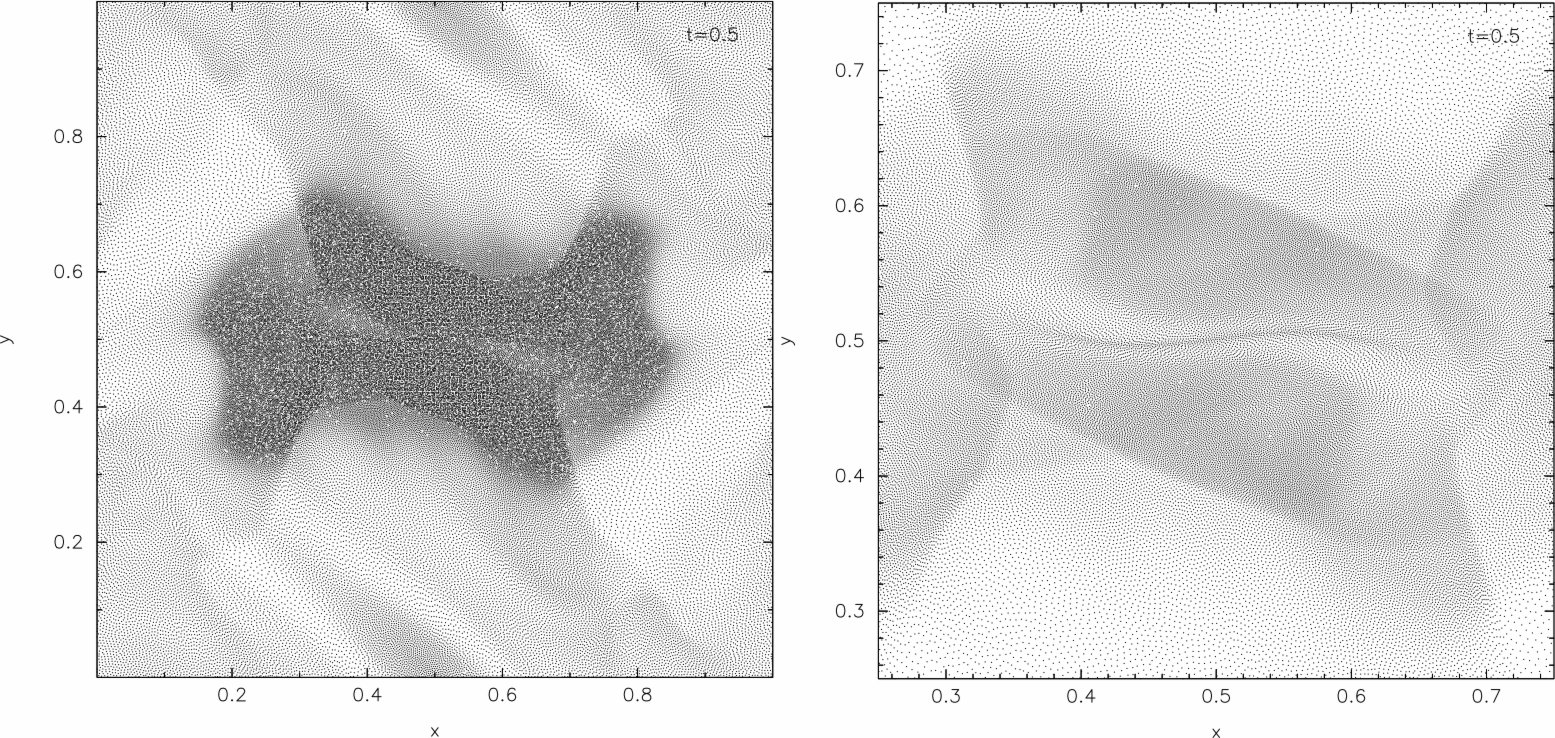}
  \caption{Particle distribution of Orszag-Tang vortex at $t = 0.5$. The left and right
    panels show the particle distribution in all domain and in the central region
    respectively. Due to Lagrangian nature of the simulation, the particle distribution
    correlated with that of the density. We would like to note that even across the
    shocks, the particle distribution remains remarkably regular without any observable
    signs of clumping.}
  \label{fig:ot5_ptcl}
\end{figure}

\begin{figure}
  \center
  \includegraphics[scale=1.0]{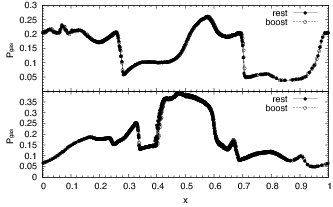}
  \caption{Pressure distribution of Orszag-Tang vortex at $t = 0.5$ and $y = 0.3125$ (top)
    and $y = 0.427$ (bottom). This can be compare to the results of other
    finite-difference or meshless MHD schemes (e.g. \citealp{2007MNRAS.379..915R,
      2008ApJS..178..137S}). In particular, our scheme has little noise in pressure, and
    the shock waves are resolve within few particles, without observable oscillatory
    behaviour in post-shock regions.}
  \label{fig:ot5_1d}
\end{figure}

In the \fig{\ref{fig:ot5_2d}} we show density at $t = 0.5$ for both rest-frame and
boost-frame initial conditions, and in \fig{\ref{fig:ot5_ptcl}} we show particle
distribution. As expected, there are no discernible differences, and both simulations
resolve discontinuities well. Furthermore, due to Lagrangian nature of the method, the
particle number density correlates with the mass density. It is worth pointing out that in
this particular simulation, the particle distribution appears to remain regular even in
the vicinity of the shocks. The \fig{\ref{fig:ot5_1d}} show 1D pressure profile at $y =
0.3125$ (top) and $y = 0.427$ (bottom) for both boosted and rest-frame simulations.  The
agreement with finite-difference scheme is excellent, ac can be compared to published
results (e.g. \citealp{2000JCoPh.161..605T}, \citealp{2007MNRAS.379..915R},
\citealp{2008ApJS..178..137S}). In contrast to shock-tube problems, no pressure blips are
visible here.

\subsection{MHD rotor}\label{sect:rotor}
\begin{figure}
  \center
  \includegraphics[scale=0.5]{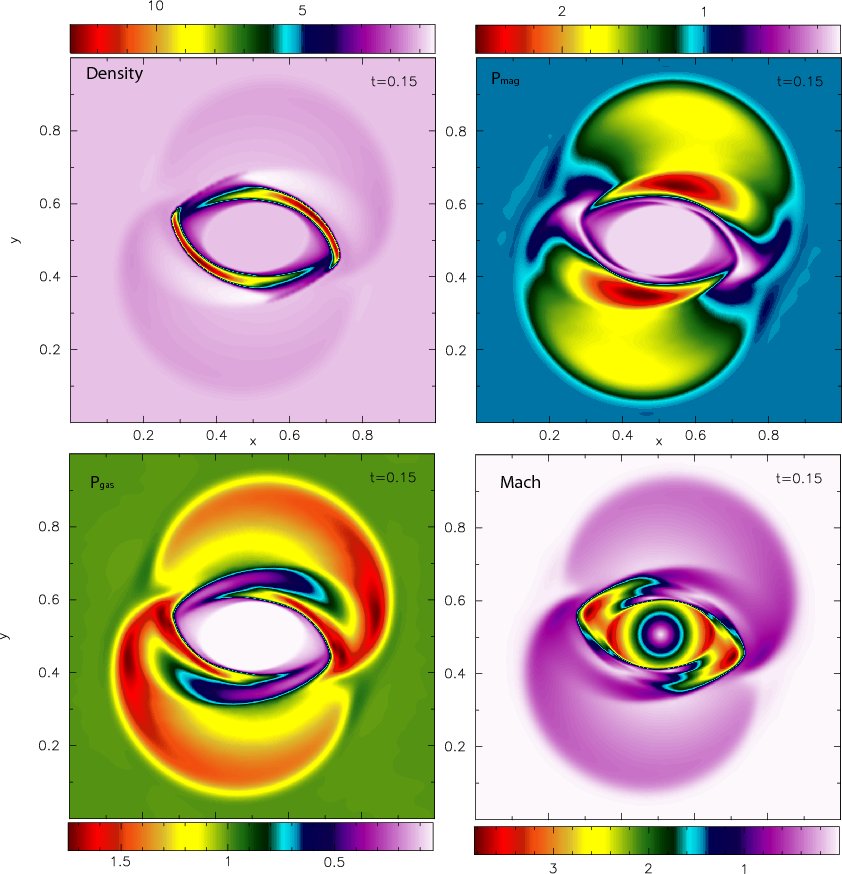}
  \caption{Density (top left), gas pressure (bottom left), magnetic pressure (top right)
    and Mach number (bottom right) plots for MHD rotor problem at $t = 0.15$. The Mach
    number panel plots Mach number using rest-frame velocities, that is first subtracting
    the boost velocity and afterwards computing mach number. The result of the rest frame
    simulation have little difference, and therefore not presented here. The colour coding
    is chosen to facilitate the comparison with published results produced by FLASH3 code
    for this problem (\citealp{2000ApJS..131..273F}, Sect 21.2.3 in FLASH 3 user guide
    manual {\tt http://flash.uchicago.edu/website/codesupport/flash3\_ug\_beta}).}
  \label{fig:rotor_2d}
\end{figure}

\begin{figure}
  \center
  \includegraphics[scale=1.0]{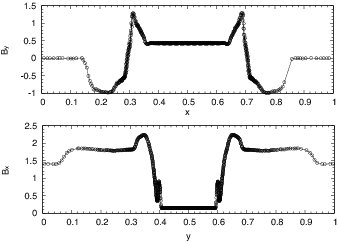}
  \caption{Distribution of $B_y$ in $y = 0.5$ slice (top pannel) and $B_x$ in $x = 0.5$
    slice (bottom) for MHD rotor problem at $t = 0.15$. This plots can be readily compared
    to Fig.26 of \citep{2008ApJS..178..137S}. The agreement is rather good everywhere,
    except for $B_y$ near $x \approx 0.34$ and $x \approx 0.66$, where the meshless scheme
    appears to diffuse the discontinuity in magnetic field.}
  \label{fig:rotor_1d}
\end{figure}

The rotor problem, introduced by \cite{1999JCoPh.149..270B} to test propagation of strong
torsional Alfv\'en waves, is also considered as one of the standard candles to validate
numerical MHD schemes. Here, the computational domain is a unit square,
$[0,1]\times[0,1]$. The initial pressure and magnetic field are uniform with values $p =
1$ and ${\bf B} = (5/\sqrt{4\pi}, 0, 0)$. Inside $R < R_0 = 0.1$ there is a dense
uniformly rotating disk with $\rho = 10$ and ${\bf v} = (-2y/R_0, +2x/R_0, 0)$, where $R =
\sqrt{x^2 + y^2}$. The ring $R_0 < R < R_1 = 0.115$ is occupied by a transition region
with $\rho = 1 + 9f(R)$ and ${\bf v} = (-2y f(R)/R, +2x f(R)/R, 0)$, where

\begin{equation}
  f(R) = \left\{
  \begin{array}{cl}
    1 & R < R_0, \\
    \frac{R_1 - R}{R_1 - R_0} & R_0 \leq R < R_1 \\
    0 &  R_1 < R.
  \end{array}
  \right.
\end{equation}
Outside $R > R_1$, the velocity is set to zero and $\rho = 1$. This problem uses an ideal
gas equation of state with $\gamma = 1.4$. The particles are distributed in the same way
as in strong blast wave problem (\S\ref{sect:blast}). The problem is solved in both rest
and boosted frames, with $v_{\rm boost} = (5/0.15, 5/0.15, 5/0.15)$. The plots of density,
gas and magnetic pressure, and mach number are shown in \fig{\ref{fig:rotor_2d}}. The 1D
slices of magnetic field at $x = 0.5$ and $y = 0.5$ are show in \fig{\ref{fig:rotor_1d}}
to facilitate comparison to Eulerian MHD schemes.

\subsection{Magneto-rotational instability}\label{sect:mri}
Magneto-rotational instability, or MRI for short, is a powerful local instability in
weakly magnetised disks that plays an important role in astrophysics
\citep{1991ApJ...376..214B, 1998RvMP...70....1B}. The ability of the scheme to model this
instability is crucial for simulation of magnetised accretion disks, or any other
simulation where non-trivial MHD effects are expected to appear. Below we conduct two
simulations that test the ability of the meshless MHD scheme to faithfully model MRI.
\subsubsection{2D axisymmetric shearing box}
The first simulation is solved in a local 2D axisymmetric shearing sheet with the initial
conditions exactly the same as in the fiducial model of \cite{2008ApJS..174..145G}. To
repeat, the unit box has a size $[0,1]\times[0,1]$ with an initially uniform density fluid
with unit density, $\rho_0 = 1$, embedded in a vertical magnetic field $B_z = B_0\sin(2\pi
x/\lambda_{\rm MRI})$, where $\lambda_{\rm MRI} = 2\pi\sqrt{16/15}v_{\rm A}/\Omega$ is the
fastest growing MRI wavelength, $v_{\rm A} = B_0/\sqrt{\rho_0}$ is Alfv\'en speed, and
$\Omega$ is angular velocity which is set to unity throughout the simulation. The magnetic
field strength $B_0 = \sqrt{2 p_0/\beta_0}$, where $\beta_0 = 1348$ in order to excite $m
= 4$ MRI mode. The instability is seeded by random velocity perturbation $\delta v = 0.01
c_s$. Isothermal gas equation of state is used, $p = c_s^2 \rho$, with $c_s = 1$ in all
domain at all times. The boundary conditions are periodic in $z$-direction, shear-periodic
$x$-direction, i.e.
\begin{equation}
  f(x,z) = f(x + n_x L_x, z + n_z L_z),
\end{equation}
\begin{equation}
  v_y(x,z) = v_y(x + n_x L_x, z + n_z L_z) + n_x q \Omega L_x.
\end{equation}
Here, $q = -1/2d\ln\Omega/d\ln R$ which is equal to $3/2$ for a Keplerian disk, and $n_x$
and $n_y$ are arbitrary integer numbers. In this shearing box model, the momentum
equation have the following form
\begin{equation}
  \partial_t(\rho{\bf v}) = -\nabla\cdot(\rho {\bf v}\otimes{\bf v} + P_T{\cal I} - {\bf
    B}\otimes{\bf B}) - 2\Omega\times(\rho{\bf v}) + 2\rho q \Omega^2 x \hat{x},
\end{equation}
where $P_T = p + B^2/2$ is the sum of the thermal and the magnetic pressure, the second
and third terms on the right hand side are Coriolis and centrifugal forces
respectively. 

\begin{figure}
  \center
  \includegraphics[scale=0.5]{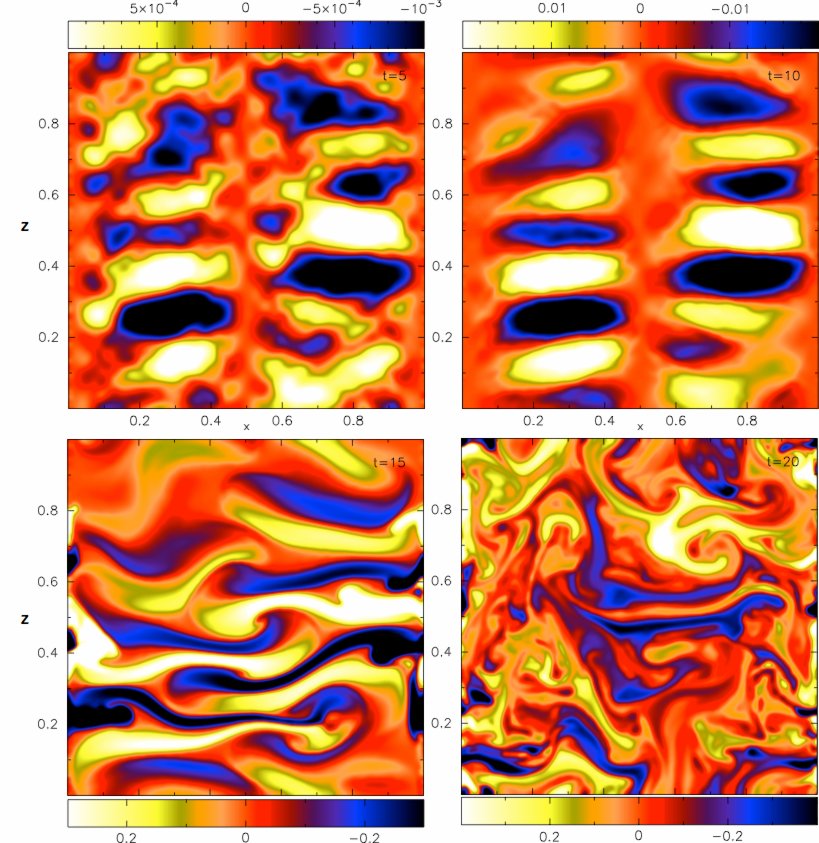}
  \caption{Toroidal, $B_y$, magnetic field at $t = 5$ (top left) , $t = 10$ (top right),
    $t = 15$ (bottom left) and $t = 20$ (bottom right) for 2D shearing box simulation at
    $256\times 256$ resolution. At $t = 10$ panel, the $m = 4$ MRI mode can be seen, in an
    excellent agreement with the MRI theory, and at $t = 20$ the flow becomes turbulent in
    agreement with finite-difference MHD schemes.}
  \label{fig:mri2d}
\end{figure}

\begin{figure}
  \center \includegraphics[scale=0.6]{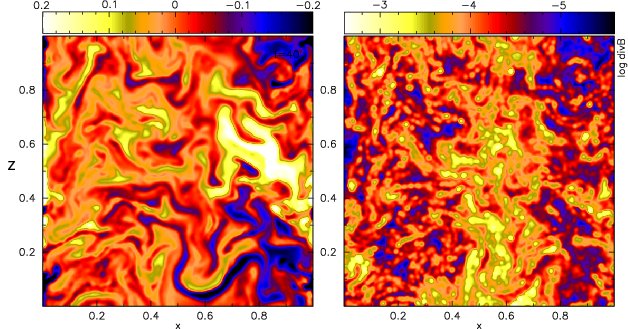}
  \caption{The figure show toroidal magnetic field at late times, $t = 40$. The aim is to
    demonstrate the ability of the scheme to keep the divergence low even in turbulent
    flows.}
  \label{fig:divB}
\end{figure}

\begin{figure}
  \center
  \includegraphics[scale=1.0]{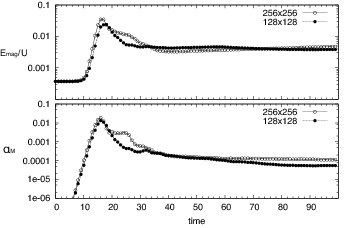}
  \caption{Time dependence of volume averaged magnetic energy and Maxwell stresses as a+
    function of time. This figure can be readily compared to Fig.4 and Fig.11 of Guan \&
    Gammie (2008). The initial growth rate is in excellent agreement with their
    results. The subsequent turbulent evolution is different, however. This is a known
    result, since the turbulent evolution of the flow depends on the numerical scheme
    employed. Due to anti-dynamo theorem, the magnetic energy and Maxwell stresses decay
    in this 2D simulation.}
  \label{fig:shear}
\end{figure}

This problem is solved in a periodic domain, with $128\times 128$ and $256\times 256$
particles initially distributed on Cartesian grid. The toroidal, $B_y$, component of
magnetic field is show in \fig{\ref{fig:mri2d}} at $t = 5$, $10$, $15$ and $20$. The $m=4$
MRI mode is clearly seen at $t = 10$, and the bottom right panel demonstrates the break up
of the laminar flow into turbulent. \fig{\ref{fig:divB}} shows both the toroidal magnetic
field (left panel) and ${\tt divB}$ (right panel) at $t=40$. The aim is to demonstrate
that even in turbulent flows, the scheme is able to keep the divergence small. Time
dependence of magnetic field energy and Maxwell-stresses is show in
\fig{\ref{fig:shear}}. The initial growth can be fit with $E_{\rm mag}/U \propto \exp(0.75
t)$, and is in excellent agreement with Fig.4 and Fig.11 of
\cite{2008ApJS..174..145G}. The decay at late times differs, which is an expected result,
since this depends on the details of numerical dissipation, which differs among MHD
schemes.

\subsubsection{3D Global disk simulation}
The final test problem studies the evolution of a circular disk around a point-mass
gravitational source. The aim is to test the particle scheme on realistic astrophysical
simulation, and to determine its weak points. The problem is set up in a computational
domain of size $[0,20]\times[0,20]\times[0,20]$ with a gravitational source of unit mass
located at origin $(10, 10, 10)$. Following methods of \cite{2004MNRAS.351..630L} and
\cite{2008ApJ...674..927A}, a circular stratified disk is sampled with $10^6$ particle
from its inner edge, $R_{\rm in} = 1$, to its outer edge, $R_{\rm out} = 4$, where $R =
\sqrt{(x-10)^2 + (y-10)^2}$ is the distance from the midplane of the disk to the
gravitational source. The scale height of the disc is $H/R = 0.1$, and only one scale
height is sampled to avoid large density variations. Particles which fall within $R_{\rm
  min} = 0.25 R_{\rm in}$ and outside the computational domain are removed from the
system. The particle distribution is regularised to minimise the start-up noise before the
initial conditions are assigned. Since this problem uses open boundary conditions, we also
imposed a lower and upper limit to the number of neighbours; specifically, we set $N_{\rm
  ngb, low} = 8$ and $N_{\rm ngb, up} = 128$ for the lower and upper limit
respectively. We did this in order to prevent particles close to the boundaries to have
either too few or too many neighbours. The lower limit, however, not been reached in our
simulations.

The initial density in the disk is $\rho(R,z) = \rho_{\rm mid} \exp(-z^2/2 H^2)$, where
$\rho_{\rm mid} = 1$ is the density in the mid plane. The scale height is $H = c_{\rm
  snd}/v_{\rm circ} R$, where $v_{\rm circ} = \sqrt{1/R}$ and $c_{\rm snd}$ is sound speed
of a particle. The latter is set to be constant in time, but has the following spatial
variation $c_{\rm snd} = (H/R) v_{\rm circ}$. Pressure is set via isothermal equation of
state, $p = c^2_{\rm snd} \rho$. The gravitational acceleration is split into horizontal
and vertical component, to make sure that the above hydrostatic equilibrium is satisfied,
namely $a_{\rm grav} = g(R) (x - 10, y - 10, z - 10)$, where $g(R) = -1/R^3$. Two
simulations where run with above initial conditions, one non-magnetised and one
magnetised. In the latter case, initial constant vertical magnetic is set with $\beta =
1348$ in the midplane, which have following $R$-variation, $B(R) = B_0 \sin(2\pi (R -
2))$, and the magnetic field outside the ring $2 < R < 3$ is set to zero. With this setup,
the orbital period of an inner disk orbit is $2\pi$. The magnetic field is forced to be
zero for $R < 1.2$ and $R > 6.5$.

\begin{figure}
  \center
  \includegraphics[scale=0.3]{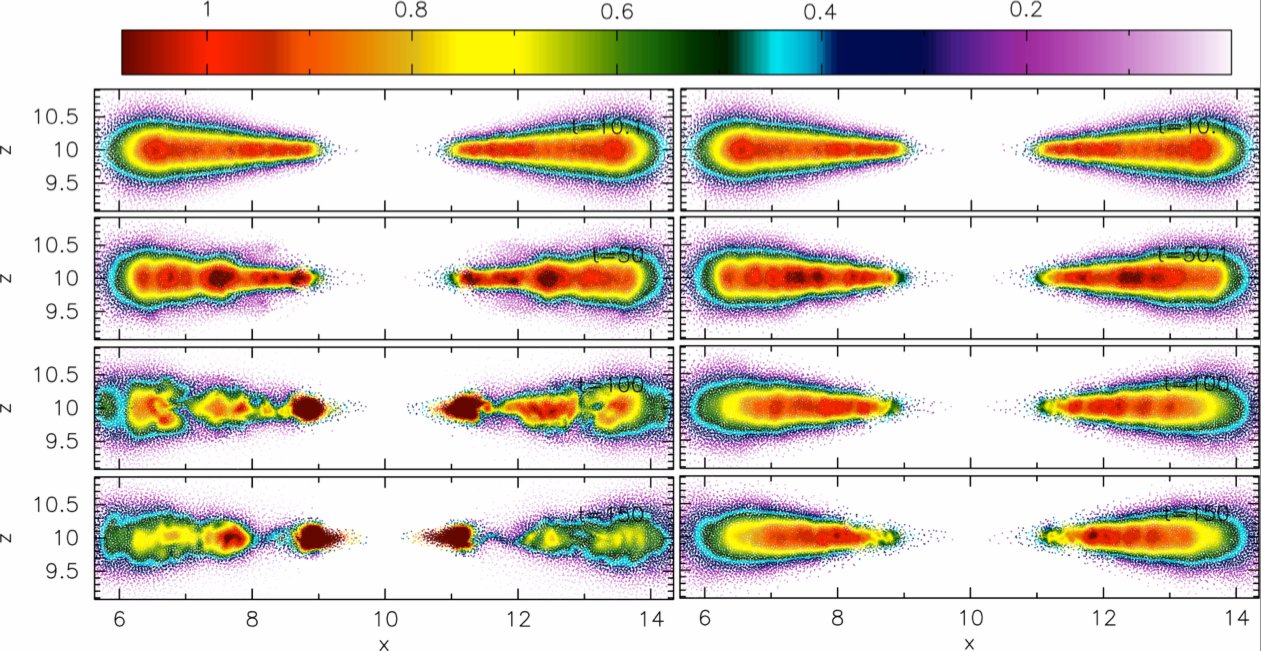}
  \caption{Density particle plot in $XZ$-plane passing through the centre $(10, 10,
    10)$. The left and right panels shows snapshot from magnetised and non-magnetised
    cases respectively at $t = 10$, $t = 50$, $t = 100$ and $t = 150$ (from top to
    bottom). The magnetic start to play an important role at $t \gtrsim 50$ or after about
    8 inner orbital periods}
  \label{fig:mri3d_rho}
\end{figure}

Density structure of the disk in $XZ$-plan passing through the gravitational source is
shown in \fig{\ref{fig:mri3d_rho}} for $t = 10$, $50$, $100$, $150$. The right panels on
the figure show density distribution for non-magnetised case. As expected, the flow
remains laminar throughout the simulation. However, at $t = 100$, the inner edge of the
disk is noticeably damaged due to particle loss through $R_{\rm min}$. This error,
generated at the inner edge of the disk, slowly propagates outward, as can be seen by the
further damage at $t = 150$. This example demonstrates the importance of proper boundary
conditions in meshless scheme. However, it is no clear how to define these. Nevertheless,
the global structure of the disk remains consistent with the initial conditions and
vertical hydrostatic equilibrium through the simulation.

\begin{figure}
  \center
  \includegraphics[scale=1.0]{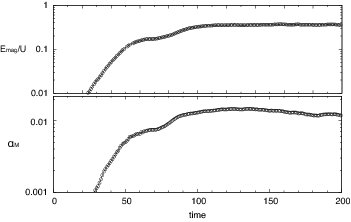}
  \caption{This figure shows the growth of the volume average magnetic field and Maxwelian
    stresses in the magnetised disk simulation. In agreement with the 2D shearing sheet
    simulation, the magnetic field and stresses reach its maximum after about 15 inner
    orbits (or 5.6 orbits at $R = 2$. In contrast to 2D case, however, the magnetic field
    does not decay furthermore, but is maintained by dynamo action.}
  \label{fig:mri3d_time}
\end{figure}

In the left panel of the \fig{\ref{fig:mri3d_rho}} we show density distribution of the
magnetised case. In contrast to non-magnetised simulation, the laminar motion breaks into
turbulence at $t \approx 100$. For earlier times, the laminar flow permits the magnetic
field to grow to $\langle \beta^{-1} \rangle \approx 0.1$ as shown in the top panel of the
\fig{\ref{fig:mri3d_time}}. Of the particular interest here, is the value of the magnetic
stresses, $\alpha_{\rm M} = -B_r B_\theta/P_{\rm gas}$, which determines the accretion
rate in the disk. Time dependence of the volume average magnetic stress is shown in the
bottom panel of the \fig{\ref{fig:mri3d_time}}. The growth continues until $t \approx
100$, after which the flow become turbulent and the volume averaged magnetic stress remain
roughly constant at $\langle \alpha_{\rm M} \rangle \approx 0.01$. In contrast to 2D
axisymmetric shearing sheet simulation, neither magnetic energy nor magnetic stressed
decay with time, implying the dynamo activity in the disk.

\begin{figure}
  \center
  \includegraphics[scale=0.5]{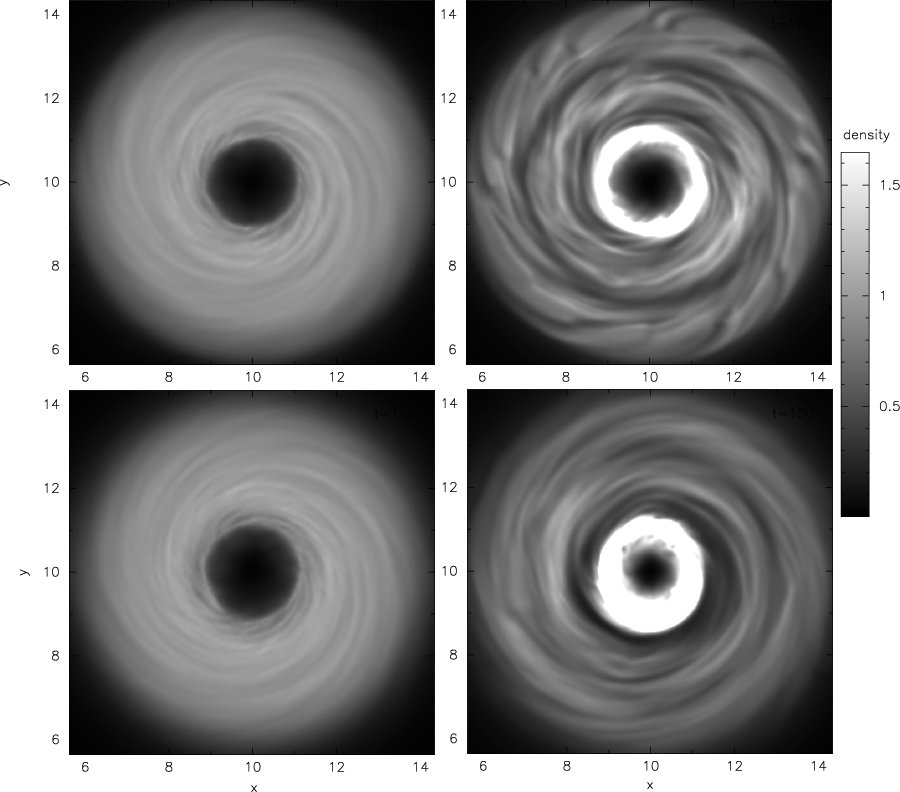}
  \caption{Density structure of the disk in $XY$-plan passing through the centre. The
    right and left panels show non-magnetised and magnetised cases respectively, at $t =
    100$ (top) and at $t = 150$ (bottom). The density structure in magnetised disk at $t =
    50$ has same form as non-magnetised, and therefore is not shown here.}
  \label{fig:mri2d_rho}
\end{figure}

\begin{figure}
  \center
  \includegraphics[scale=0.5]{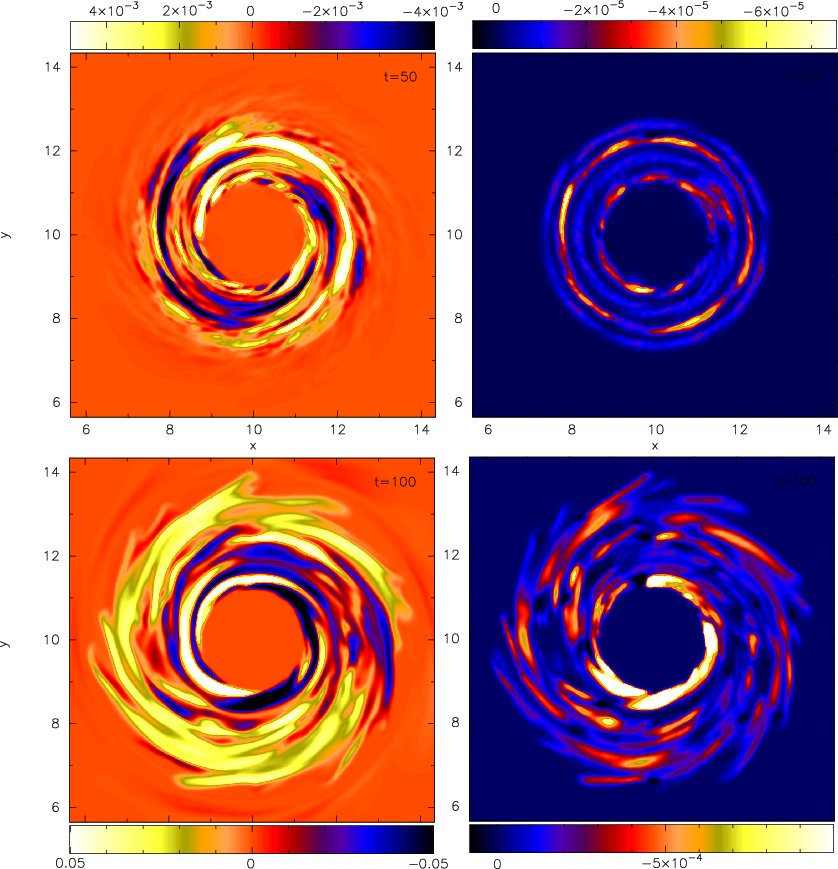}
  \caption{The left panel show the amplitude of toroidal field component, and the right
    panel the magnitude of Maxwell stress, $B_r B_\theta$ at $t = 50$ (top) and $t = 100$
    (bottom). The field and the stresses at $t = 150$ have the structures similar to those
    at $t = 100$, and therefore are not shown here.}
  \label{fig:mri2d_brbt}
\end{figure}

The density distribution in the mid plane of the disk is shown in
\fig{\ref{fig:mri2d_rho}}. The left and right panels of the figure display non-magnetised
and magnetised cases respectively, and the top and bottom panel show the profile at $t =
100$ and $t = 150$. The turbulent structure of the magnetised disk for $t > 100$ is
apparent through the existence of small scale structures in density, whereas the
non-magnetised disk remains laminar, with the exception of small spiral waves which are
caused by the error propagating from the inner disk boundary. The magnetic field structure
is shown in \fig{\ref{fig:mri2d_brbt}} for $t = 50$ (top panel) and $t = 100$ (bottom
panel). The left panel shows the amplitude of the toroidal magnetic field, $B_\theta$, and
the value of Maxwell stresses, $B_r B_\theta$. At $t = 50$, the structure begins to
appear, but the magnetic field is weak to substantially influence the dynamical evolution
of the disk. At $t = 100$, however, the magnetic field is amplified to a large values via
MRI mechanism, and has non-negligible effect on dynamics. Furthermore, the toroidal
magnetic field exhibits reverse, cause by the shear, in agreement with the theoretical
expectations.

Overall, the growth of magnetic field is consistent with published 2D and 3D shearing box
simulation. Namely, the linear MRI regime is able to amplify magnetic field to $\langle
\beta^{-1} \rangle \gtrsim 0.1$ until it breaks up into turbulence.  Due to limited
numerical resolution of this simulation, we were unable to resolve fastest growing MRI
mode, which explains the slower than expected growth. Nevertheless, the magnetic stresses
are roughly ten percent of magnetic energy, $\langle \alpha_{\rm M} \rangle \gtrsim 0.01$,
and drive the accretion of the matter. This matter is accumulate at $R < 1.25$ where
magnetic field is set to zero at boundary condition, and explains dens blob of matter in
the two lowest left panels of \fig{\ref{fig:mri3d_rho}}.

\section{Discussion and Conclusions}\label{sect:discussion}

This paper presents application of a new weighted particle scheme for conservation laws to
the equations of ideal hydrodynamics and magnetohydrodynamics. This scheme has no free
parameters which control the physics of the inter-particle interaction. There only
free-parameter in our scheme is the average number of neighbours that a particle interacts
with, and this depends only on the number of spatial dimensions. The interaction between
particles is entirely described by the source terms and fluxes. The latter are given by
the solution of the associated Riemann problem, which correctly treats dissipative
processes and discontinuous solution without explicit use of artificial viscosity or
resistivity. Due to use of Riemann solvers and reconstruction methods, our weighted
particle method is expected to be more dissipative compared to pure Lagrangian SPH without
any explicit diffusion terms. However, further quantitative comparison to SPH is required
to verify this claim in realistic astrophysical problems, where dissipative processes,
albeit locally, must be used.

In our weighted particle scheme, the smoothing length is a property of the particle
distribution only, and not the underlying solution. As a result, high resolution is
obtained in regions with high particle density, which does not need to coincide with high
mass density regions. This therefore permits similar resolution in both low and high
density regions, if the scheme is combined with particle refinement methods. In our
scheme, the physical meaning of the particle as a fluid element is lost, and the particles
should be considered as interpolation points only. Even without the refinement, the mass
of the particle can change in the course of simulations, though these changes are small in
smooth flows. The advection of a scalar field in our scheme is not as trivial as it is in
SPH. Namely, for every scalar, a transport equations must be solved which further
increase, albeit little, both memory and performance footprint of the simulation. If one
needs to follow multiple fluid composition, the matter is a bit more complicated, since
one will be required to use a consistent multi-fluid advection for chemical species
\citep{1999A&A...342..179P}, which in turn further complicates the scheme compared to
SPH. Nevertheless, we feel that the advantages of the scheme outweigh these disadvantages.

Our scheme, in principle, permits adaptive particle splitting to increase resolution in
the desired regions. However, if not done carefully, this may break the regularity of the
particle distribution, and therefore introduce substantial errors in flux divergence. The
magnitude of these errors and their impact on the outcome of the simulation are hard to
estimate. The MHD test problems with initially non-uniform but properly relaxed particle
distribution (\S\ref{sect:blast},\ref{sect:orszag} and \ref{sect:rotor}) showed excellent
results with initially nested and relaxed particle distribution. In principle, if a group
of particles is added in the course of simulation and their neighbours are adjusted to
maintain regular distribution, the noise should be small. However, further research is
required to discover optimal way for particle refinement.

We showed that the application of the weighted meshless scheme to the equations of ideal
hydrodynamics is straightforward. We also expect that such scheme is computationally
slightly more expensive than SPH. In its optimal form, it requires two loops over
neighbours, compared to one in SPH: calculation and limit of the gradients of primitive
fluid variables, and the interaction part. This is in addition to
Eq.(\ref{eq:constraint_h}), which, similarly to SPH, is solved iteratively. The
interaction part of the scheme requires solution of a Riemann problem between a particle
and its neighbours; on average, ~32 Riemann problems are solved for each particle in 3D.
Nevertheless, both HLL or HLLC solvers are only moderately expensive compared to
calculation of artificial viscosity and conductivity in SPH. However, our scheme has
higher memory footprint due to storage of gradients in the memory. Nevertheless, combined
with the lower number of neighbours usually uses in 3D SPH simulations and a large step
size in the vicinity of strong shock, which in SPH is limited by artificial viscosity, the
performance impact is minimal.

The strength of the weighted particle scheme becomes clear with its successful application
to the equations of the ideal MHD. Here, the main problem is the maintenance of
$\nabla\cdot{\bf B} = 0$ constraint. While it is not clear how to maintain this to machine
accuracy, if possible at all, in a meshless scheme, this work demonstrated that it is
possible to keep the divergence under the control by applying hyperbolic-parabolic
divergence cleaning method designed for Godunov MHD scheme. This MHD formulation is not
limited to mess-less schemes, but can also be applied to unstructured grid or moving mesh
schemes (Gaburov \& Levin 2010, in preparation), in which constraint transport
discretisation of the induction equation might be difficult or impossible to formulate.


Finally, we report that in our implementation, the science rate of this meshless MHD
scheme is roughly $10^4$ particle/s in 3D and $2\cdot 10^4$ particles/s in 2D on a single
2.7 GHz Core i7 processor core. Being a particle-based scheme, it can also be implemented
in OpenCL to allow efficient execution on many-core chips, such as GPUs. The science rate
of a GPU code which we developed and used for the simulations in this paper is $10^5$
particles/s in 3D, and twice that in 2D on GT200 chip. These are the lower values than we
expected, and with further tuning and optimisation of the code these rate can be, at
least, doubled.

\appendix

\section{Piecewise parabolic reconstruction}\label{sect:ppm_reconstruction}
Approximation of $q(x)$ in the neighbourhood of a particle $i$ is given by a second-order
Taylor expansion from the point $x_i$
\begin{equation}
  q(x) = q_i + (x - x_i)^\alpha q_i^\alpha + (x - x_i)^\alpha(x - x_i)^\beta q^{\alpha\beta}_i.
\end{equation}
To complete the reconstruction, the coefficients $q^\alpha_i$ and $q^{\alpha\beta}_i$ must
be determined. Since the number of the neighbouring particles is larger than the number of
unknown parameters, this can be done in the least-square sense by minimising the following
functional \citep{2003ApJ...595..564M}
\begin{equation}
  {\cal L}_i = \sum_j w_i \left(
  \delta q_{ij} - q^\mu_i\xi_{ij}^\mu  - q^{\mu\nu}_i\xi_{ij}^\mu\xi_{ij}^\nu
  \right)^2.
\end{equation}
Here, $\delta q_{ij} = q_j - q_i$, $\xi^\alpha_{ij} = x_j^\alpha - x_i^\alpha$ and $w_j$
is the weight of a particle $j$, which can be to $w_j = w(x_j)$ although other choices are
possible. The conditions $\partial{\cal L}_i/\partial q^\alpha_i = 0$ and $\partial{\cal
  L}_i/\partial q^{\alpha\beta}_i = 0$ result in the following set of equations
\begin{equation}
  Q^\alpha_i         =  q_i^\mu S^{\alpha\mu}_i      + q_i^{\mu\nu}S^{\alpha\mu\nu}_i ,
\end{equation}
\vspace{-0.5cm}
\begin{equation}
  Q^{\alpha\beta}_i  =  q_i^\mu S^{\alpha\beta\mu}_i + q_i^{\mu\nu}S^{\alpha\beta\mu\nu}_i .
\end{equation}
Here, $Q_i^{\alpha[\beta]} = \sum_j w_j \delta q_{ij} \xi^\alpha_{ij} [\xi^\beta_{ij}]$
and $S_i^{\alpha\beta[\mu][\nu]} = \sum_j w_j \xi^\alpha_{ij}
\xi^\beta_{ij}[\xi^\mu_{ij}]]\xi^\nu_{ij}]$, where the expressions within $[..]$ can be
    omitted to obtain $Q^\alpha$, $S^{\alpha\beta}$ and $S^{\alpha\beta\mu}$.
This system can be solved using methods of linear-algebra, by noticing that both
$(q^\alpha_i, q^{\alpha\beta}_i)$ and $(Q^\alpha_i, Q^{\alpha\beta}_i)$ coefficients can
be combined into 9-dimensional vector, and $S$ coefficients into $9\times 9$ matrix.
Finally, a parabolic reconstruction of an $i$-particle state at $x_{ij}$ is
\begin{equation}
  q_{ij;i} = q_i + \tau_i \left[ (x_{ij} - x_i)^\alpha q^\alpha_i + (x_{ij} - x_i)^\alpha
    (x_{ij} - x_i)^\beta q^{\alpha\beta}_i \right].
\end{equation}
Here, $\tau_i$ is a limiter function defined in the same way as for the linear
reconstruction (\S\ref{sect:linear_reconstruction}), with one exception: $\tau_i = 0$, if
an extrema occurs within half-vector connecting particles $i$ and $j$. With such
restriction the reconstruction is guaranteed to be monotonic.

This parabolic reconstruction is computationally more expensive compared to the linear one
due to large amount of the storage and the number of operation required to compute $Q$ and
$S$ coefficients, as well as to invert $9\times 9$ matrix. While there is advantage of
using parabolic reconstruction for Eulerian calculation, at appears to result in little
improvement when particles move with fluid velocity.

\section{HLL-type MHD Riemann solvers}\label{sect:appendix}
The one-dimensional MHD equations have the following conservative form
\begin{equation}
  \frac{\partial {\cal U}}{\partial t} + \frac{\partial{\cal G}}{\partial x} = 0,
\end{equation}
where ${\cal G} = {\cal F} - a_x {\cal U}$ is a flux in moving frame, ${\cal U}$ is a
fluid state in conservative variables and ${\cal F}$ is a flux in lab-frame:
\begin{equation}
  {\cal U} = \left(
  \begin{array}{c}
    \rho \\ 
    e \\
    \rho v_x \\
    \rho v_y \\
    \rho v_z  \\
    B_x  \\
    B_y  \\
    B_z  \\
  \end{array}
  \right), \quad
  {\cal F} = \left(
  \begin{array}{c}
    \rho v_x \\
    (e + P_T) v_x - ({\bf v}\cdot{\bf B}) B_x\\
    \rho v_x^2 + P_T - B_x^2\\
    \rho v_y v_x - B_y B_x \\
    \rho v_z v_x - B_z B_x \\
    0 \\
    B_y v_x - B_x v_y \\
    B_z v_x - B_x v_z \\
  \end{array}
  \right).
\end{equation}
Here, $P_T = p_{\rm th} + B^2/2$ is the sum of the thermal and magnetic pressures, $e =
\rho v^2/2 + e_{\rm th} + B^2/2$ is the total energy density, and $B_x$ is a constant
field, implied by $\nabla\cdot{\bf B} = 0$ constraint in 1D. The initial conditions to
this Riemann problem are given by specifying left, ${\cal U}_L$, and right, ${\cal U}_R$,
states at the interface at zero time. The flux ${\cal G}$ at any time through and normal
to the interface is given by the Riemann solver. Below, I provide formulae for two
commonly used Riemann solvers: HLL and HLLD. Both Riemann solvers reduces to the
hydrodynamic case if the magnetic field strength is zero. Namely, HLL reduces to
hydrodynamic variant of HLL solver, and HLLD reduces to HLLC presented in
\S\ref{sect:ideal_hd}.

\subsection{HLL Riemann solver}
The HLL Rieman solver use a single state to approximate intermediate wave structure
\citep{harten:35}. As a result, it is a rather diffusive solver which defuses contact
discontinuity even at rest. Nevertheless, it is computationally inexpensive and robust
Riemann solver, which can be used in pathological cases where other solvers fail. 

\begin{figure}
  \center \includegraphics[scale=0.3]{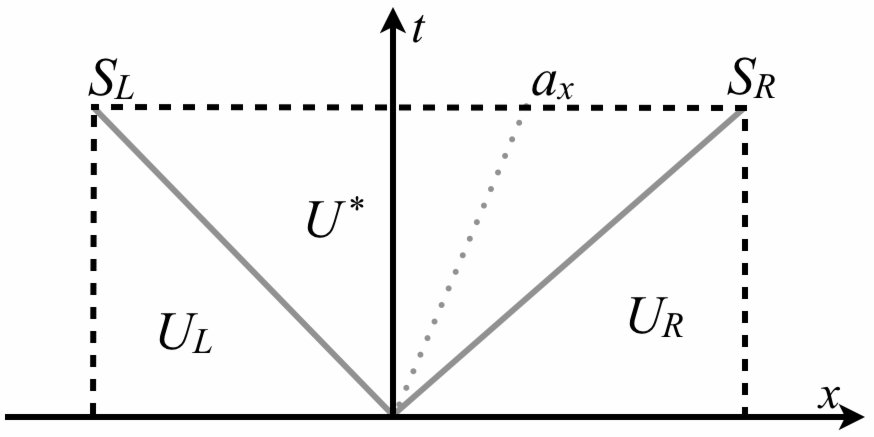}
  \label{fig:HLLE}
  \caption{This figure show wave-structure of HLL Riemann solver.}
\end{figure}

        The wave structure of the HLL solver is shown in Fig.(\ref{fig:HLLE}), and the fluxes are
given by the following expression
\begin{equation}
  {\cal G}^{HLL} = \left\{
  \begin{array}{ll}
    {\cal F}_L - a_x {\cal U}_L, & a_x < S_L, \\
    {\cal F}^\star  - a_x {\cal U}^\star, & S_L \leq a_x \leq S_R, \\
    {\cal F}_R - a_x {\cal U}_R, & S_R < S_R,
  \end{array}
  \right.
\end{equation}
where ${\cal F}_L = {\cal F}({\cal U}_L)$, ${\cal F}_R = {\cal F}({\cal U}_R)$ and
\begin{equation}
  {\cal F}^\star = \frac{S_R {\cal F}_L  - S_L {\cal F}_R + S_R S_L ({\cal U}_R - {\cal U}_L)}{S_R - S_L}
\end{equation}
is $HLL$-flux is the rest frame. The intermediate state is given by the formula
\begin{equation}
  {\cal U}^\star = \frac{S_R {\cal U}_R - S_L {\cal U}_L  - {\cal F}_R + {\cal F}_L} {S_R - S_L}.
\end{equation}
The wave speeds are $S_L = \min(v_{xL}, v_{xR}) - c_{\rm s}$ and $S_R = \max(v_{xL},
v_{xR}) + c_{\rm s}$, where $c_{\rm s} = \max(c_{{\rm s} L}, c_{{\rm s} R})$ is the
maximal signal of the left or right states. More accurate estimates based on Roe-averages
are able to reduce diffusion, but the degree of this reduction is rather small. If
diffusion needs to be minimised, one should rather use intrinsically less-diffusive
solver, such as linearised Roe-solver, or HLLD Riemann solver described next.

\subsection{HLLD Riemann solver}

\begin{figure}
  \center \includegraphics[scale=0.3]{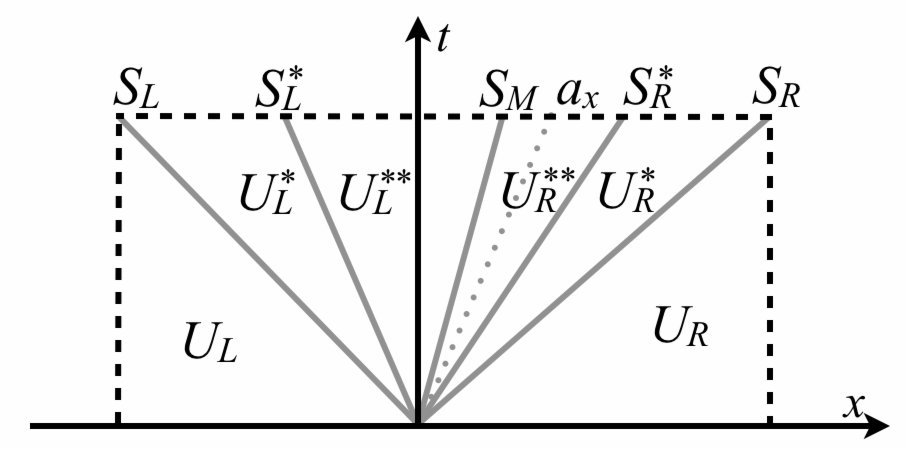}
  \label{fig:HLLD}
  \caption{This figure show wave-structure of HLLD Riemann solver.}
\end{figure}
From the wave-structure of HLLD solver (Fig.\ref{fig:HLLD}), it is clear that contact, Alfv\'en
and fast magnetosonic waves are resolved. The fluxes at the interface for this solver are
given with the following formulae
\begin{equation}
  {\cal G}^{HLLD} = \left\{
  \begin{array}{ll}
    {\cal F}_L - a_x {\cal U}_L,                                                                  & a_x < S_L, \\
    {\cal F}_L + (S_L - a_x){\cal U}^\star_L - S_L {\cal U}_L,                                           & S_L \leq a_x \leq S^\star_L, \\
    {\cal F}_L + (S^\star_L - a_x){\cal U}^{\star\star}_L - (S^\star_L - S_L){\cal U}^\star_L - S_L {\cal U}_L, & S^\star_L \leq a_x \leq S_M, \\
    {\cal F}_R + (S^\star_R - a_x){\cal U}^{\star\star}_R - (S^\star_R - S_R){\cal U}^\star_R - S_R {\cal U}_R, & S_M \leq a_x \leq S^\star_R,  \\
    {\cal F}_R + (S_R - a_x){\cal U}^\star_R - S_R {\cal U}_R,                                           & S^\star_R \leq a_x \leq S_R, \\ 
    {\cal F}_R - a_x {\cal U}_R,                                                                  & S_R < a_x.
  \end{array}
  \right.
\end{equation}
The wave speeds are $S_L = \min(v_{xL}, v_{xR}) - c_{\rm s}$ and $S_R = \max(v_{xL},
v_{xR}) + c_{\rm s}$, where $c_{\rm s} = \max(c_{{\rm s} L}, c_{{\rm s} R})$ is the
maximal signal speed in left or right state. The speed of the middle wave is defined by
\begin{equation}
  S_M = \frac{P_{T,R} - P_{T,L} + \rho v_{xL} (S_L - v_{xL}) - \rho v_{xR}(S_R - v_{xR})}
  {\rho_L(S_L - v_{xL}) - \rho_R(S_R - v_{xR})},
\end{equation}
$v^{\star\star}_{xK} = v^\star_{xK} = S_M$ for $K= L$, $R$, and $S^\star_L = S_M -
|B_x|/\sqrt{\rho^\star_L}$ and $S^\star_R = S_M - |B_x|/\sqrt{\rho^\star_R}$. The density
of both $\star$- and $\star\star$-states are
\begin{equation}
 \rho^{\star\star}_K = \rho^\star_K = \rho_K\frac{S_K - v_{xK}}{S_K - S_M},
\end{equation}
for $K = L, R$. The next four $\star$-states are
\begin{equation}
v^\star_{[yz]K} = v_{[yz]} - B_x B_{[yz]K} \frac{S_M - v_{xK}}{\rho_K(S_K - v_{xK})(S_K - S_M) - B_x^2},
\end{equation}
\begin{equation}
  B^\star_{[yz]K} = B_{[yz]K} \frac{\rho_K(S_K - v_{xk})^2}{\rho_K(S_K - v_{xK})(S_K - S_M) - B_x^2},
\end{equation}
where in $[yz]$ means either $y$- or $z$-component of the corresponding 3D
field. Following \cite{2005JCoPh.208..315M}, if the last terms on the RHS result in $0/0$
uncertainty, $v^\star_{[yzK]} = v_{[yz]K}$ and $B^\star_{[yz]K} = 0$, since in this case
there is no shock across $S_K$. Finally, the remaining two $\star$-states are
\begin{equation}
  e^\star_K = \frac{(S_K - v_{xK})e_K - P_{T_K} v_{xK} + P^\star_{T_K}S_M + B_x ({\bf
      v}_K\cdot {\bf B}_K - {\bf v}^\star_K\cdot {\bf B}^\star_K )}{S_K - S_M}.
\end{equation}
\begin{equation}
  P^\star_T = \frac{(S_R - v_{xR})\rho_R p_{T_L} - (S_L - v_{xL})\rho_L P_{T_R} +
    \rho_L\rho_R(S_R - v_{xR})(S_L - v_{xL})(v_{xR} - v_{xL})}{(S_R - u_R)\rho_R - (S_L - u_L)\rho_L},
\end{equation}
and $P^{\star\star}_{T_K} = P^\star_{T_K} = P^\star_T$, for $K = L$ , $R$. The
$\star\star$-states are
\begin{equation}
  v^{\star\star}_{[yz]} =
  \frac{\sqrt{\rho^\star_L}v^\star_{[yz]L} + \sqrt{\rho^\star_R}v^\star_{[yz]R} + (B^\star_{[yz]R} - B^\star_{[yz]L}){\rm sign}(B_x)}
       { \sqrt{\rho^\star_L} + \sqrt{\rho^\star_R} },
\end{equation}
\begin{equation}
  B^{\star\star}_{[yz]} = \frac{ \sqrt{\rho^\star_L}B_{[yz]R} +
    \sqrt{\rho^\star_R}B_{[yz]L} + \sqrt{\rho^\star_L\rho^\star_R}(v_{[yz]R} - v_{[yz]L}){\rm sign}(B_x)}
  { \sqrt{\rho^\star_L} + \sqrt{\rho^\star_R} }.
\end{equation}
\begin{equation}
  e^{\star\star}_K = e^\star_K \mp \sqrt{\rho^\star_L} ({\bf v}^\star_K\cdot{\bf B}^\star_K
  - {\bf v}^{\star\star}\cdot{\bf B}^{\star \star}){\rm sign}(B_x).
\end{equation}
In the last equations, the $-$ and $+$ on the right hand side correspond to $K= L$ and $R$
respectively. Finally, $v^{\star\star}_{[yz]K} = v^{\star\star}_{[yz]}$ and
$B^{\star\star}_{[yz]K} = B^{\star\star}_{[yz]}$.

This completes the description of HLLD Riemann solver. It is clear, this solver is
computationally more expensive than HLL solver due to its ability to also resolve contact
and Alfv\'en waves, which is important in many MHD problems. In practice, it is proven to be
a robust solver for many problems, and tests demonstrate it is at least as accurate as
linearised-Roe solver. Furthermore, HLLD solver is possible to use with equation of state
other than of ideal gas, since the equation of state in solver is used implicitly by
specifying both the gas pressure and total energy.

\section*{Acknowledgements}

This work is supported by the NWO VIDI grant \#639.042.607. The plots in this paper were
produced using SPLASH, a publicly available visualisation tool for SPH
\citep{2007PASA...24..159P}. Special thank goes to Yuri Levin and Anders Johansen for
their encouragement, patience and numerous discussions that helped to produce this work.
The authors also thank Tsuyoshi Hamada and Nagasaki University, where part of this work
has been done, for their hospitality and permission to use DEGIMA GPU-cluster.

\bibliographystyle{mn2e} \bibliography{G2010mhd}

\begin{thebibliography}{}

\bibitem[\protect\citeauthoryear{{Agertz}, {Moore}, {Stadel}, {Potter},
  {Miniati}, {Read}, {Mayer}, {Gawryszczak}, {Kravtsov}, {Nordlund}, {Pearce},
  {Quilis}, {Rudd}, {Springel}, {Stone}, {Tasker}, {Teyssier}, {Wadsley} \&
  {Walder}}{{Agertz} et~al.}{2007}]{2007MNRAS.380..963A}
{Agertz} O.,  {Moore} B.,  {Stadel} J.,  {Potter} D.,  {Miniati} F.,  {Read}
  J.,  {Mayer} L.,  {Gawryszczak} A.,  {Kravtsov} A.,  {Nordlund} {\AA}.,
  {Pearce} F.,  {Quilis} V.,  {Rudd} D.,  {Springel} V.,  {Stone} J.,  {Tasker}
  E.,  {Teyssier} R.,  {Wadsley} J.,    {Walder} R.,  2007, \mnras, 380, 963

\bibitem[\protect\citeauthoryear{{Alexander}, {Armitage}, {Cuadra} \&
  {Begelman}}{{Alexander} et~al.}{2008}]{2008ApJ...674..927A}
{Alexander} R.~D.,  {Armitage} P.~J.,  {Cuadra} J.,    {Begelman} M.~C.,  2008,
  \apj, 674, 927

\bibitem[\protect\citeauthoryear{{Balbus} \& {Hawley}}{{Balbus} \&
  {Hawley}}{1991}]{1991ApJ...376..214B}
{Balbus} S.~A.,  {Hawley} J.~F.,  1991, \apj, 376, 214

\bibitem[\protect\citeauthoryear{{Balbus} \& {Hawley}}{{Balbus} \&
  {Hawley}}{1998}]{1998RvMP...70....1B}
{Balbus} S.~A.,  {Hawley} J.~F.,  1998, Reviews of Modern Physics, 70, 1

\bibitem[\protect\citeauthoryear{{Balsara}}{{Balsara}}{2004}]{2004ApJS..151..1%
49B}
{Balsara} D.~S.,  2004, \apjs, 151, 149

\bibitem[\protect\citeauthoryear{{Balsara} \& {Spicer}}{{Balsara} \&
  {Spicer}}{1999}]{1999JCoPh.149..270B}
{Balsara} D.~S.,  {Spicer} D.~S.,  1999, Journal of Computational Physics, 149,
  270

\bibitem[\protect\citeauthoryear{{B{\o}rve}, {Omang} \& {Trulsen}}{{B{\o}rve}
  et~al.}{2001}]{2001ApJ...561...82B}
{B{\o}rve} S.,  {Omang} M.,    {Trulsen} J.,  2001, \apj, 561, 82

\bibitem[\protect\citeauthoryear{{B{\o}rve}, {Omang} \& {Trulsen}}{{B{\o}rve}
  et~al.}{2006}]{2006ApJ...652.1306B}
{B{\o}rve} S.,  {Omang} M.,    {Trulsen} J.,  2006, \apj, 652, 1306

\bibitem[\protect\citeauthoryear{{Brandenburg}}{{Brandenburg}}{2010}]{2010MNRA%
S.401..347B}
{Brandenburg} A.,  2010, \mnras, 401, 347

\bibitem[\protect\citeauthoryear{{Brio} \& {Wu}}{{Brio} \&
  {Wu}}{1988}]{1988JCoPh..75..400B}
{Brio} M.,  {Wu} C.~C.,  1988, Journal of Computational Physics, 75, 400

\bibitem[\protect\citeauthoryear{{Cha}, {Inutsuka} \& {Nayakshin}}{{Cha}
  et~al.}{2010}]{2010MNRAS.403.1165C}
{Cha} S.,  {Inutsuka} S.,    {Nayakshin} S.,  2010, \mnras, 403, 1165

\bibitem[\protect\citeauthoryear{{Cha} \& {Whitworth}}{{Cha} \&
  {Whitworth}}{2003}]{2003MNRAS.340...73C}
{Cha} S.,  {Whitworth} A.~P.,  2003, \mnras, 340, 73

\bibitem[\protect\citeauthoryear{{Colella}}{{Colella}}{1990}]{1990JCoPh..87..1%
71C}
{Colella} P.,  1990, Journal of Computational Physics, 87, 171

\bibitem[\protect\citeauthoryear{{Dedner}, {Kemm}, {Kr{\"o}ner}, {Munz},
  {Schnitzer} \& {Wesenberg}}{{Dedner} et~al.}{2002}]{2002JCoPh.175..645D}
{Dedner} A.,  {Kemm} F.,  {Kr{\"o}ner} D.,  {Munz} C.,  {Schnitzer} T.,
  {Wesenberg} M.,  2002, Journal of Computational Physics, 175, 645

\bibitem[\protect\citeauthoryear{{Dolag} \& {Stasyszyn}}{{Dolag} \&
  {Stasyszyn}}{2009}]{2009MNRAS.398.1678D}
{Dolag} K.,  {Stasyszyn} F.,  2009, \mnras, 398, 1678

\bibitem[\protect\citeauthoryear{{Evans} \& {Hawley}}{{Evans} \&
  {Hawley}}{1988}]{1988ApJ...332..659E}
{Evans} C.~R.,  {Hawley} J.~F.,  1988, \apj, 332, 659

\bibitem[\protect\citeauthoryear{{Fryxell}, {Olson}, {Ricker}, {Timmes},
  {Zingale}, {Lamb}, {MacNeice}, {Rosner}, {Truran} \& {Tufo}}{{Fryxell}
  et~al.}{2000}]{2000ApJS..131..273F}
{Fryxell} B.,  {Olson} K.,  {Ricker} P.,  {Timmes} F.~X.,  {Zingale} M.,
  {Lamb} D.~Q.,  {MacNeice} P.,  {Rosner} R.,  {Truran} J.~W.,    {Tufo} H.,
  2000, \apjs, 131, 273

\bibitem[\protect\citeauthoryear{{Gottlieb} \& {Shu}}{{Gottlieb} \&
  {Shu}}{1998}]{Gottlieb98totalvariation}
{Gottlieb} S.,  {Shu} C.-W.,  1998, Math. Comp, 67, 73

\bibitem[\protect\citeauthoryear{{Guan} \& {Gammie}}{{Guan} \&
  {Gammie}}{2008}]{2008ApJS..174..145G}
{Guan} X.,  {Gammie} C.~F.,  2008, \apjs, 174, 145

\bibitem[\protect\citeauthoryear{Harten, Lax \& van Leer}{Harten
  et~al.}{1983}]{harten:35}
Harten A.,  Lax P.~D.,    van Leer B.,  1983, SIAM Review, 25, 35

\bibitem[\protect\citeauthoryear{{Inutsuka}}{{Inutsuka}}{2002}]{2002JCoPh.179.%
.238I}
{Inutsuka} S.,  2002, Journal of Computational Physics, 179, 238

\bibitem[\protect\citeauthoryear{Lanson \& Vila}{Lanson \&
  Vila}{2008a}]{1404789}
Lanson N.,  Vila J.-P.,  2008a, SIAM J. Numer. Anal., 46, 1912

\bibitem[\protect\citeauthoryear{Lanson \& Vila}{Lanson \&
  Vila}{2008b}]{1404790}
Lanson N.,  Vila J.-P.,  2008b, SIAM J. Numer. Anal., 46, 1935

\bibitem[\protect\citeauthoryear{{Lodato} \& {Rice}}{{Lodato} \&
  {Rice}}{2004}]{2004MNRAS.351..630L}
{Lodato} G.,  {Rice} W.~K.~M.,  2004, \mnras, 351, 630

\bibitem[\protect\citeauthoryear{{Maron}}{{Maron}}{2005}]{2005prpl.conf.8461M}
{Maron} J.,  2005, in Protostars and Planets V {Gradient Particle
  Magnetohydrodynamics (GPM), a Lagrangian Particle Algorithm with Fourth-Order
  Gradients and Magnetic Fields}.
pp 8461--+

\bibitem[\protect\citeauthoryear{{Maron} \& {Howes}}{{Maron} \&
  {Howes}}{2003}]{2003ApJ...595..564M}
{Maron} J.~L.,  {Howes} G.~G.,  2003, \apj, 595, 564

\bibitem[\protect\citeauthoryear{{Mignone} \& {Tzeferacos}}{{Mignone} \&
  {Tzeferacos}}{2010}]{2010JCoPh.229.2117M}
{Mignone} A.,  {Tzeferacos} P.,  2010, Journal of Computational Physics, 229,
  2117

\bibitem[\protect\citeauthoryear{{Miyoshi} \& {Kusano}}{{Miyoshi} \&
  {Kusano}}{2005}]{2005JCoPh.208..315M}
{Miyoshi} T.,  {Kusano} K.,  2005, Journal of Computational Physics, 208, 315

\bibitem[\protect\citeauthoryear{{Monaghan}}{{Monaghan}}{2000}]{2000JCoPh.159.%
.290M}
{Monaghan} J.~J.,  2000, Journal of Computational Physics, 159, 290

\bibitem[\protect\citeauthoryear{{Monaghan}}{{Monaghan}}{2002}]{2002MNRAS.335.%
.843M}
{Monaghan} J.~J.,  2002, \mnras, 335, 843

\bibitem[\protect\citeauthoryear{Monaghan}{Monaghan}{2005}]{0034-4885-68-8-R01}
Monaghan J.~J.,  2005, Reports on Progress in Physics, 68, 1703

\bibitem[\protect\citeauthoryear{{Orszag} \& {Tang}}{{Orszag} \&
  {Tang}}{1979}]{1979JFM....90..129O}
{Orszag} S.~A.,  {Tang} C.,  1979, Journal of Fluid Mechanics, 90, 129

\bibitem[\protect\citeauthoryear{{Plewa} \& {M{\"u}ller}}{{Plewa} \&
  {M{\"u}ller}}{1999}]{1999A&A...342..179P}
{Plewa} T.,  {M{\"u}ller} E.,  1999, \aap, 342, 179

\bibitem[\protect\citeauthoryear{{Powell}, {Roe}, {Linde}, {Gombosi} \& {de
  Zeeuw}}{{Powell} et~al.}{1999}]{1999JCoPh.154..284P}
{Powell} K.~G.,  {Roe} P.~L.,  {Linde} T.~J.,  {Gombosi} T.~I.,    {de Zeeuw}
  D.~L.,  1999, Journal of Computational Physics, 154, 284

\bibitem[\protect\citeauthoryear{{Press}, {Teukolsky}, {Vetterling} \&
  {Flannery}}{{Press} et~al.}{1992}]{1992nrca.book.....P}
{Press} W.~H.,  {Teukolsky} S.~A.,  {Vetterling} W.~T.,    {Flannery} B.~P.,
  1992, {Numerical recipes in C. The art of scientific computing}

\bibitem[\protect\citeauthoryear{{Price}}{{Price}}{2007}]{2007PASA...24..159P}
{Price} D.~J.,  2007, Publications of the Astronomical Society of Australia,
  24, 159

\bibitem[\protect\citeauthoryear{{Price}}{{Price}}{2010}]{2010MNRAS.401.1475P}
{Price} D.~J.,  2010, \mnras, 401, 1475

\bibitem[\protect\citeauthoryear{{Price} \& {Bate}}{{Price} \&
  {Bate}}{2008}]{2008MNRAS.385.1820P}
{Price} D.~J.,  {Bate} M.~R.,  2008, \mnras, 385, 1820

\bibitem[\protect\citeauthoryear{{Price} \& {Monaghan}}{{Price} \&
  {Monaghan}}{2004}]{2004MNRAS.348..123P}
{Price} D.~J.,  {Monaghan} J.~J.,  2004, \mnras, 348, 123

\bibitem[\protect\citeauthoryear{{Price} \& {Monaghan}}{{Price} \&
  {Monaghan}}{2005}]{2005MNRAS.364..384P}
{Price} D.~J.,  {Monaghan} J.~J.,  2005, \mnras, 364, 384

\bibitem[\protect\citeauthoryear{{Price} \& {Rosswog}}{{Price} \&
  {Rosswog}}{2006}]{2006Sci...312..719P}
{Price} D.~J.,  {Rosswog} S.,  2006, Science, 312, 719

\bibitem[\protect\citeauthoryear{{Rosswog} \& {Price}}{{Rosswog} \&
  {Price}}{2007}]{2007MNRAS.379..915R}
{Rosswog} S.,  {Price} D.,  2007, \mnras, 379, 915

\bibitem[\protect\citeauthoryear{{Springel} \& {Hernquist}}{{Springel} \&
  {Hernquist}}{2002}]{2002MNRAS.333..649S}
{Springel} V.,  {Hernquist} L.,  2002, \mnras, 333, 649

\bibitem[\protect\citeauthoryear{{Stone}, {Gardiner}, {Teuben}, {Hawley} \&
  {Simon}}{{Stone} et~al.}{2008}]{2008ApJS..178..137S}
{Stone} J.~M.,  {Gardiner} T.~A.,  {Teuben} P.,  {Hawley} J.~F.,    {Simon}
  J.~B.,  2008, \apjs, 178, 137

\bibitem[\protect\citeauthoryear{Swegle, Hicks \& Attaway}{Swegle
  et~al.}{1995}]{Swegle1995123}
Swegle J.~W.,  Hicks D.~L.,    Attaway S.~W.,  1995, Journal of Computational
  Physics, 116, 123

\bibitem[\protect\citeauthoryear{Toro}{Toro}{1999}]{toro99}
Toro E.~F.,  1999, Riemann Solvers and Numerical Methods for Fluid Dynamics : A
  Pratical Introduction.
Springer, 2nd edition

\bibitem[\protect\citeauthoryear{{T{\'o}th}}{{T{\'o}th}}{2000}]{2000JCoPh.161.%
.605T}
{T{\'o}th} G.,  2000, Journal of Computational Physics, 161, 605

\bibitem[\protect\citeauthoryear{van Leer}{van Leer}{2006}]{citeulike:3115494}
van Leer B.,  2006, Communications in Computational Physics, 1, 192

\bibitem[\protect\citeauthoryear{Vila}{Vila}{1999}]{VILA:1999}
Vila J.~P.,  1999, Mathematical models and Methods in Applied Sciences, 9, 161

\end{thebibliography}

\end{document}